\documentclass[aps,prb,floatfix,twocolumn,nopacs,longbibliography,superscriptaddress]{revtex4-2}
\usepackage{comment}
\usepackage[utf8]{inputenc}
\usepackage[american]{babel}
\usepackage[T1]{fontenc}
\usepackage[pdftex]{graphicx}  
\usepackage{graphicx, xcolor}
\usepackage{nicefrac}
\usepackage{dcolumn}
\usepackage{bm}
\usepackage{amsmath,amsthm,amssymb}
\usepackage{amsfonts}
\usepackage{bbold}
\usepackage{hyperref}
\usepackage{xcolor}
\usepackage[normalem]{ulem}
\hypersetup{colorlinks,bookmarksopen,bookmarksnumbered,
citecolor=blue,
linkcolor=blue,
pdfstartview=false,
urlcolor=blue}
\usepackage{graphicx}
\usepackage{wrapfig}
\usepackage{soul}
\usepackage{mathtools}
\usepackage{float}
\usepackage{natbib}

\newcommand{\m}[1]{\mathcal{#1}}

\newcommand{\antp}[1]{\ifmmode\textcolor{blue}{\text{#1}}\else\textcolor{blue}{#1}\fi}

\begin{document}

\global\long\def\ket#1{\left|#1\right\rangle }%
\global\long\def\bra#1{\left\langle #1\right|}%

\global\long\def\braket#1#2{\left\langle#1\mid#2\right\rangle }%
\global\long\def\expected#1{\left\langle #1\right\rangle }%

\title{Heating Dynamics of Correlated Fermions under Dephasing}

\author{Antonio Picano}
\affiliation{JEIP, UAR 3573 CNRS, Coll\`ege de France,   PSL  Research  University, 11,  place  Marcelin  Berthelot,75231 Paris Cedex 05, France}
\author{Matthieu Vanhoecke}
\affiliation{JEIP, UAR 3573 CNRS, Coll\`ege de France,   PSL  Research  University, 11,  place  Marcelin  Berthelot,75231 Paris Cedex 05, France}
\author{Marco Schir\`o}
\affiliation{JEIP, UAR 3573 CNRS, Coll\`ege de France,   PSL  Research  University, 11,  place  Marcelin  Berthelot,75231 Paris Cedex 05, France}

\begin{abstract}
We study the dissipative dynamics of correlated fermions evolving in presence of a local dephasing bath. To this extent we consider the infinite coordination limit of the corresponding Lindblad master equation, provided by Dynamical Mean-Field Theory for open quantum systems. We solve the resulting quantum impurity problem, describing an Anderson impurity coupled to a local dephasing, using weak-coupling perturbation theory in interaction and dephasing. We show that the dissipative dynamics describes heating towards infinite temperature, with a relaxation rate that depends strongly on interaction. The resulting steady-state spectral functions are however non-trivial and show an interplay between coherent quasiparticle peak and local dephasing. We then discuss how thermalization towards infinite temperature emerges within DMFT, by solving the impurity problem throughout its self-consistency. We show that thermalization under open quantum system dynamics is qualitatively different from the closed system case. In particular, the thermalization front found in the unitary is strongly modified,  a signature of the irreversibility of the open system dynamics. 
\end{abstract}

\date{\today}
\maketitle

\section{Introduction}
% \textcolor{red}{To do List}
% \begin{enumerate}
%     \item \textcolor{red}{May be add a short summary how is defined $\Gamma(t,t^\prime)$ ? Because Photoexcitation is not usual in the Lindblad/open quantum system field no ? in the paper should be adress both for the DMFT  and the Lindblad community no ? }
%     \item \textcolor{red}{short discussion on the motivation of the photo-excitation ?} 
% \end{enumerate}
Open quantum many-body systems where coherent evolution competes with dissipative dynamics due to coupling to external environments are currently attracting widespread interest~\cite{fazio2025manybodyopenquantumsystems,sieberer2025universality}.  A popular framework to describe dissipation which is relevant for several experimental platforms across atomic physics, solid state and quantum optics, is the Lindblad master equation which describes Markovian open quantum system dynamics~\cite{breuerPetruccione2010}.

Among the different dissipative mechanisms, dephasing due to energy exchange with a hot bath has been studied extensively~\cite{esposito2005exactly,esposito2005emergence,znidaric2010exact,eisler2011crossover,znidaric2013transport,cai2013barthel,medvedyeva2016exact,turkeshi2021diffusion,alba2021spreading,jin2022exact,wellnitz2022rise,catalano2023anomalous,zn9v-k73w}. Several works have focused on the dissipation induced heating and the effect of many-body interactions on the heating dynamics~\cite{gerbier2010heating,pichler2010nonequilibrium,sarkar2014light,yanay2014heating}. In particular, for bosonic systems, this has been triggered by pioneering theoretical and experimental works~\cite{poletti2012interaction,poletti2013emergence,bouganne2020anomalous,bernier2018lightcone,vatre2023dynamics}.  The heating dynamics of many-body interacting fermionic systems has been less explored, even though few results are available systems~\cite{buchhold2015nonequilibrium,bernier2020melting,bacsi2020vaporization,tindall2019heating,bernier2014dissipative}. This topic is particularly relevant to both quantum simulation experiments~\cite{daley2022practical} %as well as in 
and
solid state physics. 
%On one hand indeed 
Indeed,
recent progress in quantum simulations of fermionic gases in optical lattices have achieved record breaking cold temperatures and started to reveal intriguing features of strong correlation physics~\cite{xu2025neutral,chalopin2024probingmagneticoriginpseudogap}. An important issue is to understand the mechanisms for heating production. In solid-state context dephasing can be seen to arise from the coupling to a hot phonon bath~\cite{korolev2024unveilingroleelectronphononscattering}.

At the more fundamental level the question of how quantum many-body system heat up is relevant for many different nonequilibrium problems. The case of dephasing is particularly appealing in this respect since one expects thermalization towards a featureless infinite temperature state. How precisely this thermalization arises in a quantum many-body system and the role of the self-consistent environment made by the rest of the system is unclear and largely unexplored. Furthermore, the connection between unitary evolution where memory of the initial condition is preserved indefinitely, and dissipative dynamics which admits a well defined steady state is worth to be explored, particularly in a regime where dissipation is parametrically weak with respect to coherent energy scales.

In this work we tackle these questions by studying the dissipative dynamics of correlated fermions, described by a Fermi-Hubbard model, evolving in presence of a local dephasing bath. To this extent we consider the infinite coordination limit of the corresponding Lindblad master equation, provided by Dynamical Mean-Field Theory (DMFT) for open quantum systems~\cite{Georges1996,aoki2014nonequilibrium,Scarlatella_2021_DMFT}. We solve the resulting quantum impurity problem, describing an Anderson impurity coupled to a local dephasing~\cite{vanhoecke2024diagrammatic,vanhoecke2025kondozeno}, using weak-coupling perturbation theory in interaction and dephasing. We show that the dissipative dynamics describes heating towards infinite temperature, with a relaxation rate that depends strongly on interaction. The resulting steady-state spectral functions are however non-trivial and show an interplay between coherent quasiparticle peak and local dephasing. We then discuss how thermalization towards infinite temperature emerges within DMFT, by solving the impurity problem throughout its self-consistency. We show that thermalization under open quantum system dynamics is qualitatively different from the closed system case. While the latter is characterized by a ballistic thermalisation front that signals the emergence of a self-consistent bath~\cite{Picano2025}, the irreversible dissipative dynamics does not retain memory of initial condition and so lack any sharp front.

The paper is structured as follows. In Sec.~\ref{sec:model} we introduce the Hubbard model and the dissipative dynamics with dephasing that will be the focus of this work. In Sec.~\ref{sec:DMFT} we present a self-contained introduction to DMFT for open quantum many-body systems and describe the impurity solver used in our work to solve DMFT equations. Sec.~\ref{sec:quench} contains our results concerning the dynamics after a quench of the dephasing, while in Sec.~\ref{sec:photoexc} we discuss the dynamics after photoexcitation and a simultaneous
quench of the dephasing. In Sec.~\ref{sec:front} we discuss and interpret the emergence of infinite temperature thermalization in DMFT. Finally, Sec.~\ref{sec:conclusion} is devoted to discussions and conclusions. Several technical Appendix complete this work with further details and results.

%%%%%%%%%%%%%%%%%%%%%%%%%%%%%%%%%%%%%%%%%%%%%%%%%%%%%%%%%%%
\begin{figure*}[t!]
    \centering
    \includegraphics[width=0.9\linewidth]{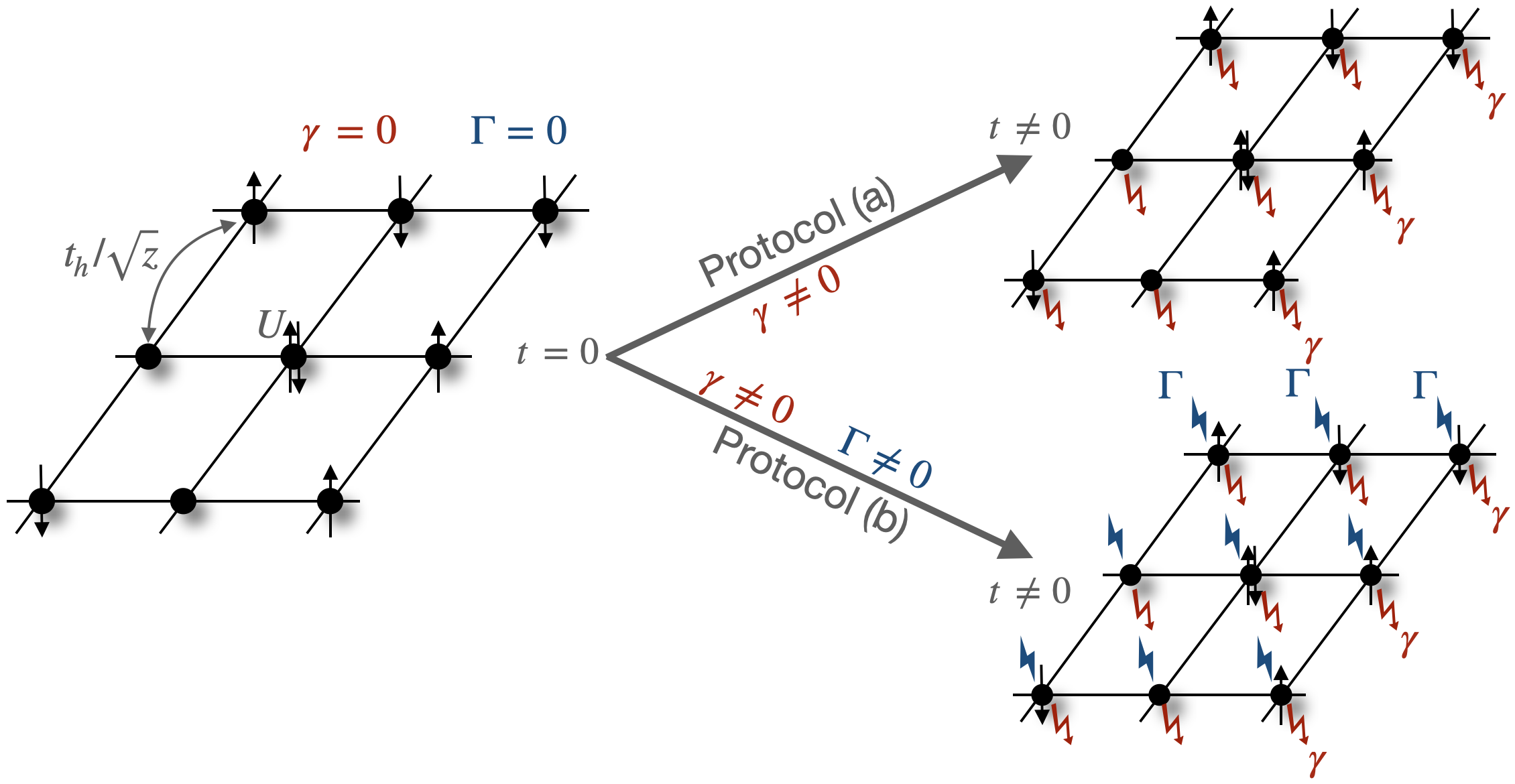}
    \caption{ Quench protocol for the dissipative Fermi-Hubbard model — An interacting lattice system, characterized by the Coulomb interaction $U$ and hopping amplitude $t_h / \sqrt{z}$ at $t=0^+$ is subjected to a quench, involving either (a) Markovian dissipation alone ($\gamma \neq 0$), or (b) both Markovian dissipation ($\gamma \neq 0$) and photoexcitation ($\Gamma \neq 0$).}
    \label{fig:Sketch_DMFT}
\end{figure*}
%%%%%%%%%%%%%%%%%%%%%%%%%%%%%%%%%%%%%%%%%%%%%%%%%%%%%%%%%%%

\section{Model and Dynamics}\label{sec:model}

We consider a system of interacting fermionic particles on a lattice with coordination number $z$, coupled to a dissipative bath (see sketch in Fig.~\ref{fig:Sketch_DMFT}). The many-body density matrix of the system $\rho_t$ evolves according to a Lindblad master equation~\cite{breuerPetruccione2010}
\begin{align}
    \label{eqn:Lindblad}
    \partial_t \rho_t = -i \left[ H,\rho_t \right] + \sum_{i ,\sigma} L_{i\sigma} \rho_t L_{i\sigma}^\dagger - \frac{1}{2} \{L_{i\sigma}^\dagger L_{i\sigma} ,\rho_t \} 
    %= \mathcal{L} \rho
\end{align}
with a set of jump operators $L_{i\sigma}$, which we assume to be local at each site, accounting for dissipative processes. The index \textit{i} corresponds to the site index,  while $\sigma$ represents the spin. 
%( in general the second index can represent any quantum number). 
The coherent evolution is governed by the Hubbard Hamiltonian which reads,
\begin{align}\label{eqn:Hubbard}
    H =- \frac{t_h}{\sqrt{z}}\sum_{\langle i,j \rangle, \sigma}   c_{i\sigma}^\dagger c_{j\sigma} + U\sum_{i} \left(n_{i\uparrow}-\frac{1}{2}\right)\left(n_{i\downarrow}-\frac{1}{2}\right)
\end{align}
where the hopping $t_h$ is restricted to next-neighboring sites, $n_{i\sigma}= c_{i\sigma}^\dagger c_{i \sigma}$ is the spin-density operator for spin $\sigma$ on site $i$, and $U$ is the Coulomb energy. 
In the following, we consider a Bethe lattice, with hopping $t_h=1$ setting the unit of energy and semi-elliptic density of states $D(\epsilon)=\sqrt{4-\epsilon^2}/(2\pi)$. For what concerns the local dissipative dynamics, we consider the case of dephasing, corresponding to jump operators of the form
\begin{align}\label{eq:jump_dephasing}
    L_{i\sigma} = \sqrt{\gamma} n_{i\sigma}
    %= \sqrt{\gamma_\sigma} d_\sigma^\dagger d_\sigma 
\end{align}
where $\gamma$ is the dephasing rate associated to the spin $\sigma$. This type of dissipation describes energy exchange with a collection of independent baths at infinite temperature. Indeed one can show that the total energy $E(t)=\mbox{Tr}\left(\rho_t H\right)$ is not conserved under the Lindblad dynamics with Eq.~(\ref{eq:jump_dephasing}). On the contrary, the total number of particle is conserved both under the unitary part of the dynamics as well as by the dissipation. Since the jump operators are Hermitian one can easily show that the trivial density matrix $\rho_{\infty}=1$, corresponding to a featureless infinite temperature state, is a steady state of the Lindblad dynamics for any finite sistem size. 

In the following we will discuss the dynamics of the system after two types of nonequilibrium protocols (see sketch in Fig.~\ref{fig:Sketch_DMFT}): (a) A sudden switching-on of the dephasing starting from an equilibrium configuration; (b) the dynamics after a photoexcitation and simultaneous quench of the dissipation, see details below. 
%In both cases we discuss the  approach to the steady state as well as the spectral functions on top ofthe steady-state which instead remain non-trivial.
In both cases, we discuss the  approach to the steady state as well as the spectral functions in the  steady-state. 
%which instead remain non-trivial.
Before presenting our results we briefly discuss the method used to solve our model.

\section{Out-of-Equilibrium DMFT for Markovian fermions}\label{sec:DMFT}

To investigate the out-of-equilibrium dynamics of the dissipative Hubbard model introduced in Sec.~\ref{sec:model}, we employ an extension of dynamical mean-field theory (DMFT) to Markovian open quantum systems~\cite{Scarlatella_2021_DMFT}. This approach allows us to nonperturbatively treat local quantum correlations while capturing the effects of both coherent evolution and Markovian dissipation. In particular, we consider the infinite coordination number limit $z \rightarrow \infty$ of the Bethe lattice, where DMFT becomes exact, and the lattice problem can be mapped onto a quantum impurity problem subject to a self-consistently determined bath.

The starting point for writing DMFT for Markovian fermions is to formulate the Lindblad master equation in the framework of non-equilibrium Keldysh field theory~\cite{kamenev2011nonequilibrium,SiebererRepProgPhys2016,stefanucci2024kadanoff}. In this formalism, the evolution of the system is encoded in the partition function $\mathcal{Z}$ defined on the Keldysh time contour $\mathcal{C}$,
\begin{align}
    \label{eqn:Z}
    \mathcal{Z} = \int \prod_{i,\sigma} \mathcal{D}\left[ c_{i\sigma},\bar{ c}_{i\sigma}\right] e^{iS}
\end{align}
where $\mathcal{S}$ is the Keldysh action.
The latter is written in terms of the fermionic coherent state fields $\{ \bar{ c}_{i\sigma\alpha}, c_{i\sigma\alpha}\}$, which are defined on the forward $(\alpha=+)$ and backward $(\alpha=-)$ branches of the Keldysh contour for each lattice site i and spin $\sigma$:
\begin{align}
	\label{eqn:S_imp}
    \mathcal{S} = \int_{-\infty}^{+\infty} dt \sum_{\alpha=\{-,+\}} \sum_{i\sigma} \alpha \bar{ c}_{i\sigma \alpha} i\partial_t c_{i\sigma \alpha} -i\int_{-\infty}^{+\infty} dt \mathcal{L}
\end{align}
The coherent and dissipative contributions to the dynamics are captured through the Lindbladian term $\mathcal{L}$, which in the Keldysh formalism becomes:
\begin{align}\label{eqn:Lindblad_Keldysh}
    \mathcal{L} = &-i  \left( H_{+ }  - H_{-}\right) + \notag \\ & + \sum_{i,\sigma} L_{i\sigma+}  \bar{ L}_{i\sigma-} - \frac{1}{2} \left( \bar{ L}_{i\sigma+}L_{i\sigma+}+\bar{ L}_{i\sigma-} L_{i\sigma-}\right)
\end{align}
In the infinite coordination limit, one can apply the cavity method: By isolating a single site (the "impurity") and integrating out the remaining lattice degrees of freedom, the influence of the surrounding sites is captured through a self-consistent bath, a frequency dependent non-Markovian environment
which effectively encodes the feedback from the rest of the system. Therefore, in the infinite coordination limit the effective action of the impurity reads
\begin{align}\label{eqn:qim}
	\mathcal{S}
	=\mathcal{S}_{\text{loc}}-i
	\int_\mathcal{C} dt dt' 
	\sum_\sigma
	\bar{c}_\sigma(t) \Delta_\sigma(t,t')c_\sigma(t')\,,
\end{align}
where $\mathcal{S}_{\text{loc}}$ includes all local contributions coming from both coherent evolution (i.e. interactions) and dissipation, while $\Delta_{\sigma}(t,t^\prime)$ is the hybridization function of the DMFT bath, which characterizes both the spectral properties and occupation of the effective bath. Within the DMFT framework, this hybridization function 
%is not fixed a priori but 
is determined self-consistently from the impurity Green's function,
\begin{align}
G_{\sigma}(t,t')=-i\langle T_{\mathcal{C}} c_{\sigma}(t)c^{\dagger}_{\sigma}(t')\rangle
\end{align}

For a Bethe lattice with semi-elliptic density of states, the DMFT self-consistency condition closes in a simple analytic form~\cite{Georges1996}, yielding to:
\begin{align}
	\label{eqn:Self_consistent_Eq_bethe}
	\Delta_\sigma(t,t')=t_h^2G_\sigma(t,t')
\end{align}
Simulating the dynamics of the original dissipative Hubbard model within DMFT reduces to solving self-consistently the impurity problem defined by Eq.~\eqref{eqn:qim}. In the present case this %reduces to
takes the form of 
a dissipative Anderson impurity model~\cite{vanhoecke2024diagrammatic,vanhoecke2025kondozeno} describing a single interacting spinful level with Coulomb repulsion and local dephasing.
%computing the corresponding Green’s function, and iterating the self-consistency condition until convergence is reached.
%We now discuss in detail the numerical approach used to solve this impurity problem and the physical observables extracted from it.
Below we detail the numerical approach used for the solution of the impurity problem and outline the physical observables extracted from it.

\subsection{Impurity Solver and DMFT Self-consistent loop}
%\subsection{Self-consistent loop}
We focus here on the impurity solver and the iterative numerical scheme for the DMFT self-consistency. We begin by defining the \textit{non-interacting} (i.e., bath-decoupled) impurity Green's function $G_{0,\sigma}(t,t')$, which obeys the Dyson equation:
    \begin{align}
	   \label{Gweiss}
	   G_{0,\sigma}^{-1}(t,t')=[i\partial_t -H_{\text{loc},\sigma}(t)]\delta_{\mathcal{C}}(t,t') \,,
    \end{align}
where $H_{\rm{loc},\sigma}=h_\sigma(t) +\epsilon_d-\mu$ denotes the local impurity Hamiltonian. It contains eventually the single-particle energy $\epsilon_d$, the chemical potential $\mu$, or the Hartree term $h_\sigma(t) = U n_\sigma(t)$ with the density $n_\sigma(t)$ per spin. In the present case we study a half-filled system, so that $\mu = U/2$, $\epsilon_d=0$ and $H_{\rm loc,\sigma}(t)  = 0 $. 

At the beginning we assume that the fully interacting impurity Green's function is simply given by $G_0$, and we calculate the hybridization $\Delta$ with Eq.~\eqref{eqn:Self_consistent_Eq_bethe}. Then, for each $(t,t')$ in the Keldysh contour we solve the self-consisitent DMFT loop:    
    
\begin{enumerate}
    \item The Weiss impurity Green's function $\mathcal{G}$ is obtained by incorporating the hybridization of the bath via:
    \begin{align}
	   \label{Gweiss}
	   \mathcal{G}^{-1}_\sigma(t,t') = G_{0,\sigma}^{-1}(t,t') - \Delta_\sigma(t,t') \,,
    \end{align}
    where $\Delta_\sigma(t,t')$ is the hybridization function obtained from the previous DMFT iteration or initial guess.

    \item The full interacting impurity Green’s function $G_\sigma(t,t')$ is computed via the Dyson equation:
    \begin{align}
	   \label{Gfull}
	   G_{\sigma}^{-1}(t,t') = \mathcal{G}^{-1}_\sigma(t,t') - \Sigma_{\text{int},\sigma}(t,t') \,,
    \end{align}
    where $\Sigma_{\text{int},\sigma}(t,t')$ is the impurity self-energy, incorporating both coherent ($\Sigma_{U ,\sigma}$) and dissipative ($\Sigma_{\gamma,\sigma}$) interaction:
    \begin{align}
		\label{Sigma_int}
	   \Sigma_{\text{int}}(t,t')
	   \equiv
	   \Sigma_{U,\sigma}(t,t')+\Sigma_{\gamma,\sigma}(t,t')
    \end{align}
    The Hubbard interaction term $\Sigma_{U,\sigma}$ is computed using second-order iterated perturbation theory (IPT):
    \begin{align}
	   \label{IPT}
	   \Sigma_{U,\sigma}(t,t') = U(t) U(t')\, \mathcal{G}_\sigma(t,t')\, \mathcal{G}_{\bar\sigma}(t',t)\, \mathcal{G}_{\bar\sigma}(t,t')\,.
    \end{align}
    The purely dissipative part is captured through an additional local self-energy term
    \begin{align}
	   \label{dephasing}
	   \Sigma_{\gamma,\sigma}(t,t') = \gamma \, G_\sigma(t,t)\, \delta(t,t') \,,
    \end{align}
    where $\gamma$ denotes is local dephasing rate (see Appendix~\ref{sec:AppendixSelfEnergyDerivation} for the derivation).

    \item The updated Green's function $G_\sigma(t,t')$ is then used to compute a new hybridization function via the DMFT self-consistency condition~\eqref{eqn:Self_consistent_Eq_bethe},
    \begin{align}
    	\label{bethe_selfcons}
    	\Delta_\sigma(t,t') = t_h^2\, G_\sigma(t,t') +\Gamma(t,t') \,,
    \end{align}
where we introduced an additional contribution to the hybridization function, $\Gamma(t,t')$, in order to simulate a photo-excitation. 
This excitation is modeled as a pulse characterized by a duration $T_0=1$ and an amplitude $\Gamma$, representing the strength of the coupling induced by the pulse. Details on the implementation of $\Gamma(t,t^\prime)$ can be found in Appendix~\ref{sec:ex_prot}.
\end{enumerate}
The procedure is iterated until self-consistency is reached within a prescribed numerical tolerance. In summary, for each time argument $(t,t')$ on the Keldysh contour, self-consistency is reached by iterating till convergence the following loop:
\begin{align}
G_\sigma^{-1}(t,t') = &[i\partial_t %+\mu
-H_{\text{loc}}(t)]\delta_{\mathcal{C}}(t,t') - t_h^2 G_\sigma(t,t')  \nonumber \\
&- \Sigma_{\text{int},\sigma}(t,t') - \Gamma(t,t') 
    \label{DMFT}
\end{align}

\begin{figure*}[t!]
    \centering
    \includegraphics[width=1.0\linewidth]{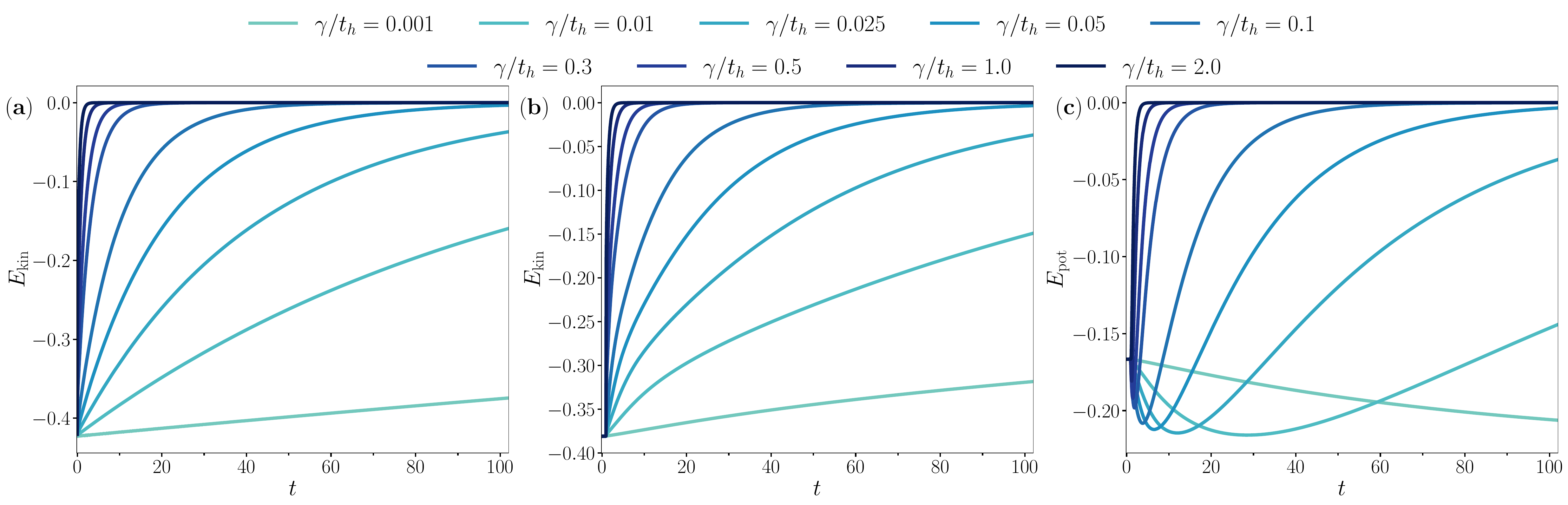}
    \caption{ Dissipative Fermi-Hubbard Model - Dynamics of the kinetic and potential energy for increasing values of $\gamma$. Panel (a) shows the kinetic energy in the non-interacting Fermi-Hubbard model, where the potential energy is zero due to the absence of coherent interactions. Panels (b) and (c) show the kinetic and potential energy, respectively, for the interacting case with $U/ t_h =2$. }
    \label{fig:fig1}
\end{figure*}
%%%%%%%%%%%%%%%%%%%%%%%%%%%%%%%%%%%%%%%%%%%%%%%%%%%%%%%%%%%%%%%%%%%%

\subsection{Quantum Boltzmann Equation}
\label{sec:QBE}
%The memory occupation for the solution of the Kadanoff-Baym Eqs.~\eqref{eqn:KBEretarded} to \eqref{eqn:KBElesser}  scales like $\mathcal{O}(n^2)$ to propagate to time $n dt$. This represents a bottleneck for long time dynamics simulations. A possible way to overtake the limitation is represented by the recently developed non-perturbative Quantum Boltzmann Equation (QBE)~\cite{Picano2021}.

When solving Dyson equations like Eq.~\eqref{DMFT}, a major task is to obtain the Green’s functions $G(t,t')$ with their dependence on two time arguments from the many-body self-energy $\Sigma(t,t')$ via the Kadanoﬀ-Baym equations (KBE). The latter are  equations of motion for $G$, in which $\Sigma$ acts as
a memory kernel. The main numerical cost is given by
the evaluation of memory integrals, i.e., the convolution
of $G$ and $\Sigma$ over earlier times. For an equidistant
time discretization with $N_t$ steps, the computational
eﬀort and the required computer memory shows a
cubic scaling $\mathcal{O}(N_t^3)$ and a quadratic scaling $\mathcal{O}(N_t^2)$
with $N_t$, respectively. 
This represents a bottleneck for studying the long-time heating dynamics of our electronic system. There are various paths to overcome this bottleneck~\cite{Stahl2022}.
Here we compute the time evolution of the system in Eq.~\eqref{DMFT}  by means of a non-perturbative QBE for the energy distribution function
\begin{align}
\label{FG}
F( \omega, t ) = \frac{\Im G^< (\omega, t)}{2 \pi \mathcal{A} (\omega,t)}
\end{align} 
solved in the framework of DMFT~\cite{Picano2021}, where $G^<$ is the lesser component of the interacting Green's function $G$ and $\mathcal{A}(\omega)=-\frac{1}{\pi} \Im G^R(\omega+i0)$ the spectrum and $t$ is the average time. 
The QBE gives an equation 
\begin{align}
\label{QBE}
\partial_t F ( \omega, t ) = I_\omega[F(\omega,t), \cdot ]
\end{align}
for the evolution of the electronic distribution function, with scattering integral:
\begin{align} 
\label{scatt_int}
%I_{\omega} [F] = & \Im{ \Sigma^<( \omega, t )}  + 2 F ( \omega, t ) \Im{ \Sigma^R (\omega, t)},
I_{\omega} [F] = &-i \{\Sigma^<(\omega,t)+\Sigma^R(\omega,t)F(\omega,t)-F(\omega,t) \Sigma^A(\omega,t) \}
\end{align}
where $\Sigma(\omega, t)=\Sigma_{\text{int}}(\omega, t)+\Delta(\omega, t)$.
% $\Sigma_{\text{int}}(\omega,t)$ and the spectrum $\mathcal{A}(\omega,t)$ in Eq.~\eqref{ness-local-dmft} are understood in terms of an auxiliary steady-state DMFT impurity model with given prescribed distribution function $\bar F(\omega)=F(\omega,t)$: 
% \begin{align}
% \label{ness-local-dmft}
% \Sigma_{\text{int}}(\omega,t)=&
% \Sigma_\omega^{\text{NESS}}[F(\cdot,t)],\,\,
% \mathcal{A}(\omega,t)=
% \mathcal{A}_\omega^{\text{NESS}}[F(\cdot,t)]
% \end{align}
% where NESS stays for non-equilibrium steady-state.
The evaluation of the functionals $\Sigma_{\text{int}}(\omega,t)$ and $\mathcal{A}(\omega,t)$ is done iteratively by solving a non-equilibrium steady-state (NESS) impurity problem with the suitable DMFT impurity solver. 
In practice, for each time $t$, $F(\omega, t+h)\equiv \bar F(\omega)$ is given by the QBE Eq.~\eqref{QBE} and 
is not updated during the NESS loop described below, while
$\Sigma_{\text{int}}(\omega,t+h) \equiv \bar \Sigma_{\text{int}}(\omega)$ 
and $\mathcal{A}(\omega,t+h) \equiv \bar{\mathcal{A}}(\omega)$ are calculated self-consistently in the following way:
\begin{enumerate}
	\label{NESSloop}
	\item 
	Start from a guess for $\bar \Sigma_{\text{int}}(\omega)$ (if you are at the very first timestep, $t=0$, take for example the equilibrium $\bar{\Sigma}_{\text{int}}$,  otherwise start from $\bar{\Sigma}_{\text{int}}(\omega)$ calculated at the previous timestep) and solve the steady-state equation for $\bar G^R(\omega)$, 
	\begin{align}
	\label{G_R}
	& \bar G^R(\omega)
	= 
	[\omega +i0^+ - H_{\text{loc}}(t) - \bar{\Delta}^R(\omega)- \bar \Sigma_{\text{int}}^R(\omega)]^{-1}.
	\end{align}
	in order to determine $ \bar{ \mathcal{A}}(\omega)=-\frac{1}{\pi} \Im \bar G^R(\omega+i0)$.
	We recall that in our case (half-filling) it is always $ H_{\text{loc}}(t) =0$. 
	\item
	Determine the lesser Green's function from the given distribution function, using the steady-state variant of the fluctuation-dissipation theorem:
	\begin{align}
	\bar G^<(\omega) = 2\pi i \bar F(\omega) \bar{\mathcal{A}}(\omega)	
	\end{align} 
	In NESS, $\bar F(\omega)$ does not need to be a Fermi-Dirac distribution and, indeed, in general it is not.
	%, and its values is given by the QBE~\eqref{QBE} that is calculated before starting the NESS loop.
	\item
	Use the self-consistency Eq.~\eqref{eqn:Self_consistent_Eq_bethe} to fix the hybridization function of the effective steady state impurity model,
	\begin{align}
	\label{Delta_QBE}
	\bar{\Delta}^{R,<}(\omega)=t_h^2\bar G^{R,<}(\omega) + \Gamma(\omega)
	\end{align}
    where $\Gamma(\omega)$ is the additional contribution to the hybridization function that implement to simulate photoexcitation, see Appendix \ref{sec:ex_prot}.
	\item
	Solve the impurity model. With IPT as impurity solver, we first determine $\bar{\mathcal{G}}(\omega)$ from $\bar{\Delta}(\omega)$,
    \begin{align}
        \bar{\mathcal{G}}^R(\omega) &= [\omega +i0^+ - H_{\text{loc}}(t) - \bar{\Delta}^R(\omega)]^{-1} \,, \nonumber \\
        \bar{\mathcal{G}}^<(\omega) &= \bar{\mathcal{G}}^R(\omega) \bar{\Delta}^<(\omega) \bar{\mathcal{G}}^A(\omega) \,, 
    \end{align}
    then Fourier transform to relative time, evaluate the steady-state version of Eq.~\eqref{IPT}, and transform back to frequency space to obtain $\bar \Sigma_U^{R,<}(\omega)$.
	\item In the time-translational invariant case (NESS loop), the purely dissipative part of the self-energy, Eq.~\eqref{dephasing}, becomes
	\begin{align}\label{eqn:KBE_SelfEnergy}
		\bar \Sigma^{R,<}_{\gamma}(\omega)=  \gamma(t) \frac{1}{2 \pi} \int d \omega \ \bar G^{R,<}(\omega)
	\end{align} 

    \item Set $\bar \Sigma_{\text{int}}(\omega) = \bar \Sigma_U(\omega)+ \bar \Sigma_\gamma(\omega)$, and iterate  steps $1)$ to $5)$ until convergence.
\end{enumerate}

The DMFT self-consistency serves as a way to evaluate $\Sigma^{\text{NESS}}[F_G(\cdot,t)]$ (as well as $\mathcal{A}^{\text{NESS}}[F_G(\cdot,t)]$ and $\Delta^{\text{NESS}}[F_G(\cdot,t)]$). 
%NESS stays for non-equilibrium steady-state.
The differential equation \eqref{QBE} is then solved using a Runge-Kutta algorithm in fourth order, and, once $F(\omega, t+h)$ is obtained, the new NESS loop at time $t+h$ is solved.  

The  total energy of the system at (average) time $t$ is
\begin{align}
E_{\text{QBE}}(t)&=\frac{1}{2 \pi}\int d\omega \left \{-2i [\Delta_\sigma(\omega,t) G_\sigma(\omega,t)]^<\right \} \nonumber \\
&+\frac{1}{2 \pi}\int d\omega \left \{ - i [ \Sigma_{U, \sigma}(\omega,t) G_\sigma(\omega,t)]^< \right \}
\label{energy_QBE}
\end{align}
The non-perturbative QBE approach
%, which relies on some assumptions (the main of which is the separation of timescales, see~\cite{Picano2021}), 
gives a computational effort which scales like $\mathcal{O}(t_{\text{max}})$, as compared to the $\mathcal{O}(t_{\text{max}}^3)$ of the Noneq-DMFT solution. 
Moreover, in QBE the memory occupied does not depend on $t_{\text{max}}$, while in Noneq-DMFT it scales like  $\mathcal{O}(t_{\text{max}}^2)$. The approximation on which the non-perturbative QBE relies is the separation of timescales, i.e., the time evolution of the system has to be slow with respect to relevant internal energy diﬀerences in the system, such as the linewidth or relevant spectral features~\cite{Picano2021}.

\section{Results: Quench of Dissipation}\label{sec:quench}

We now present our results for the dissipative dynamics of the Fermi-Hubbard model with dephasing, solved in the framework of the non-perturbative QBE with DMFT. The system is initialized at time $t=0^{-}$ in an interacting thermal state at temperature $\beta_i =20 $. At time $t=0^+$,  the dephasing rate $\gamma$ is suddenly switched on, initiating the out-of-equilibrium dynamics. 
%\textcolor{red}{only dephasing is switched on, interaction is there in the initial state as well right?}
%\antp{yes, the initial temperature is $\beta_i=20$ in all the simulations, and $U$ is unchanged before and after $t=0$, so the quench is only in $\gamma$.}
%For the initial state we consider a semi-circular density of states $D(\omega) = \sqrt{4-\omega^2}/(2 \pi)$ corresponding to a Bethe lattice with hopping amplitude $t_h=1$, which sets the unit of energy, and its inverse defines the unit of time. 
%AP: I comment out the previous entence because we already mentioned before equation (3) that we are considering a bethe lattice with semielliptic dos

\subsection{Heating dynamics}
We start by briefly discussing the dissipative but non-interacting case, $U=0$, corresponding to a dissipative free fermion lattice system. We stress that due to the dephasing the system is still interacting, i.e non-gaussian in the dissipative sense. However the specific nature of the dissipation makes it possible to write the Green's function into a Dyson equation for the impurity solver (see Appendix~\ref{sec:AppendixNoninteractingCase} for more details). In Fig.~\ref{fig:fig1} (a), we show the time evolution of the kinetic energy for increasing values of $\gamma$. We see that the kinetic energy increases exponentially towards zero, $E_{\rm kin}=0$, at long times for all values of $\gamma$.  This exponential scaling is essentially characterized by a decay constant set by the dissipation rate. The steady-state value $E_{\rm kin}=0$ is compatible with an infinite temperature state, as we will confirm later by looking at the distribution function and the effective temperature.

%%%%%%%%%%%%%%%%%%%%%%%%%%%%%%%%%%%%%%%%%%%%%%%%%%%%%%%%%%%%%%%%%%%%
\begin{figure*}[t!]
    \centering
    \includegraphics[width=1.0\linewidth]{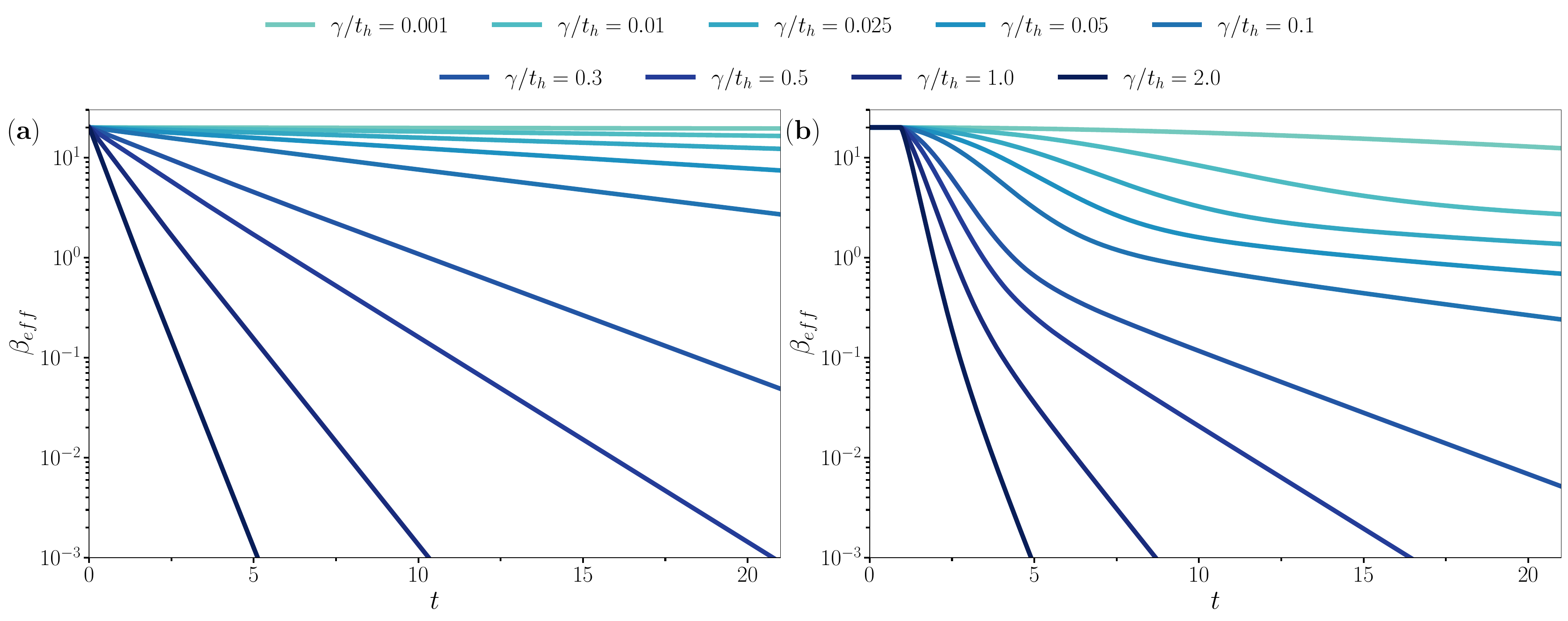}
    \caption{ Dissipative Fermi-Hubbard Model - Dynamics of the effective temperature $\beta_{\rm eff}$ for increasing values of dephasing $\gamma$. Panel (a) shows the non-interacting case $U/t_h = 0$, while panel (b) displays the interacting case with $U/t_h = 2$. The initial temperature of the system is $\beta_i=20$. 
    }
    \label{fig:Quench_gamma_Effective_temperature}
\end{figure*}
%%%%%%%%%%%%%%%%%%%%%%%%%%%%%%%%%%%%%%%%%%%%%%%%%%%%%%%%%%%%%%%%%%%%

We now turn to the fully interacting regime and analyze the heating dynamics in the presence of local dephasing. Our goal is to highlight the role of interactions in shaping the relaxation behavior of the system. As shown in Fig.~\ref{fig:fig1}(b), the dynamics of the kinetic energy still reaches the infinite temperature regime for $U/t_h=2$, however the thermalization timescale is strongly influenced by the interaction strength $U$ and the dissipation rate $\gamma$. This 
%sensitivity
becomes particularly 
%apparent
evident
in the weakly dissipative regime, where we see %$E_{\rm kin}$ first grows more rapidly at short-times and then slow-down.
the kinetic energy firstly growing more rapidly at short times, and then slowing down for longer times.
This behavior can be understood more clearly by examining the potential energy dynamics in Fig.~\ref{fig:fig1}(c). We observe an initial decrease in the potential energy, followed by a minimum at a characteristic time $t^\ast$, after which it begins to rise. The early-time decrease reflects the interplay between coherent interactions and dissipation, where the system initially tends to minimize its energy under the influence of $U$ before dissipation becomes dominant. This intermediate regime, governed by the competition between coherent and dissipative processes, corresponds to the time window where the system exhibits a metastable prethermal plateau 
(see also Fig.~\ref{fig:Quench_gamma_Effective_temperature}(b) and the subsequent related discussion).
However, at later times, dissipation begins to dominate and drives the system towards an infinite-temperature state, resulting in a monotonic increase in the potential energy as a function of time. The time $t^\ast$ thus marks the crossover from interaction–dissipation dynamics to a purely dissipation-driven regime. While $t^\ast$ scales approximately as $t^\ast \propto 1/\gamma$ and remains almost independent of the interaction strength U,  the potential energy at the crossover point $E_{\rm pot}(t^\ast)$ (not shown), decreases with increasing U, which leads to a more pronounced prethermal plateau at larger interaction strengths.

\subsection{Effective Temperature Dynamics}

The emergence of infinite temperature thermalization is more clearly and unambiguously seen in the dynamics of the distribution function $F(\omega,t)$ and in particular from its low-frequency component, which reflects the onset of an effective temperature. We extract a time-dependent effective temperature as  
$F(\omega,t)\sim -\beta_{\rm eff}(t)\omega/4$ around $\omega=0$
%$\partial F(\omega,t)|_{\omega=0}\sim \beta_{\rm eff}(t)$ 
and plot $\beta_{\rm eff}$ as a function of time for different values of dephasing in Fig.~\ref{fig:Quench_gamma_Effective_temperature}. In absence of interactions, panel (a), the inverse effective temperature vanishes exponentially in time with a rate that grows with $\gamma$. For $U/t_h=2$ instead, as shown in panel (b), we see a different behavior. The dynamics of $\beta_{\rm eff}$ is at first faster, at short-times, due to interaction effects, and then slows down developing a prethermal plateau for weak values of dissipation. In particular,
for $\gamma / t_h < 0.025$, the dynamics exhibits clear signatures of prethermalization: 
%where 
The effective temperature reaches a plateau and remains approximately constant over an extended time window, even as the total energy continues to evolve slowly. 
%This decoupling of effective temperature and energy signals the emergence of a long-lived prethermal state, which becomes more pronounced with increasing interaction strength, as $U$ becomes larger and the prethermal plateaus become more long-lived and defined.
This decoupling of effective temperature and energy indicates the emergence of a long-lived prethermal state, which becomes more pronounced with increasing interaction strength, as larger values of 
$U$ lead to longer-lived and more well-defined prethermal plateaus.
For larger values of $\gamma$, dissipation quickly overcomes the effects of coherent interactions, making the system rapidly escape the prethermal plateau and relax exponentially fast toward the infinite-temperature limit. 

%%%%%%%%%%%%%%%%%%%%%%%%%%%%%%%%%%%%%%%%%%%%%%%%%%%%%%%%%%%%%%%%%%%%

\begin{figure}[t!]
    \centering    \includegraphics[width=1\linewidth]%{figures_paper/Figure2_U_3.pdf}
    {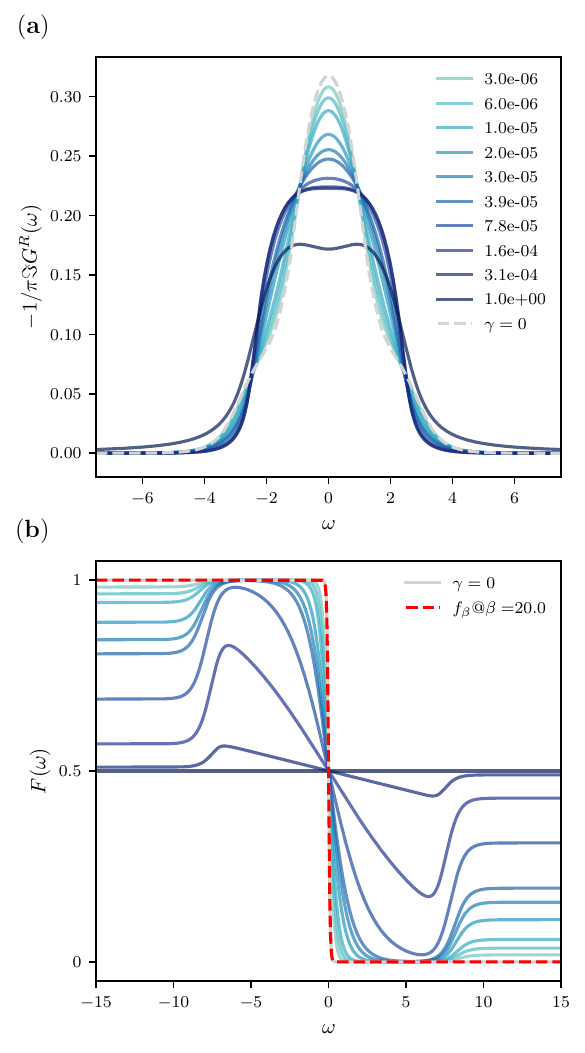}    
    \caption{Dissipative Fermi-Hubbard Model - Spectral properties and distribution function of the interacting Fermi-Hubbard model ($U/t_h= 2$) for increasing values of dephasing $\gamma$ at $t=12500$. Panel (a): Spectral function in the prethermal steady-state as a function of the dissipative rate $\gamma$. Panel (b): The corresponding distribution function, with the red dotted line showing the equilibrium fermi function at $\beta_i=20$ and $\gamma=0$.}
\label{fig:particle_density_spectra}
\end{figure}
%%%%%%%%%%%%%%%%%%%%%%%%%%%%%%%%%%%%%%%%%%%%%%%%%%%%%%%%%%%%%%%%%%%%

%This is further confirmed by looking at the time-dependent effective temperature, defined from the distribution function, see Fig.~\ref{fig:Quench_gamma_Effective_temperature}.

%This is further confirmed by looking at the time-dependent effective temperature, defined from the distribution function.  We see the latter decays exponentially to zero, with a decay constant which is essentially set by the dissipation rate. 

\subsection{Prethermal Steady-State Spectrum and Occupation}

We then move to the steady-state which we characterize using the nonequilibrium Green's functions, in particular the spectrum and the occupation encoded in the retarded Green's function and in the distribution function. For moderate  to high values of dephasing, the system approaches a genuine thermal state at infinite temperature and as a result the distribution function becomes completely flat in frequency, see Fig.~\ref{fig:particle_density_spectra}.
%The spectral function remains non-trivial, however the coherent quasiparticle peak is destroyed by dissipation and only broad incoherent features at high energy, corresponding to Hubbard bands remain visible in the spectrum. 
The spectral function remains non-trivial; however, the coherent quasiparticle peak is destroyed by dissipation, and only broad, incoherent high-energy features corresponding to Hubbard bands remain visible in the spectrum. 
The situation at weak dephasing is, %instead,
however, more interesting. We see in Fig.~\ref{fig:particle_density_spectra} that the prethermal steady-state features a highly non-thermal occupation. At low frequency we can still extract an effective temperature, %which matches quantitatively with the data of the dynamics. 
as we did for the data in Fig.~\ref{fig:Quench_gamma_Effective_temperature}. Yet, we see that finite frequency excitations are not suppressed uniformly towards the infinite temperature state: in particular, the modes around the edge of the single particle bandwidth are the slowest to heat up.
%However excitations at finite frequencies, around the edge of the single particle bandwidth are the slowest to heat up and remain 
%Upon further increasing the dephasing we see that the weight of the distribution function at positive and negative frequencies decreases, even if slowly around $\omega\sim \pm U/2$.
%\antp{(not sure I fully understood this last sentence)}. 

Looking at the spectral function we see that the prethermal state looks like a correlated Fermi Liquid, with a local quasiparticle peak that is robust to weak dissipation and that melts down as $\gamma$ increases. The transfer of spectral weight from low to high frequencies leads to an increase of the Hubbard bands, as already discussed.

\section{Results: Photoexcitation}\label{sec:photoexc}

\begin{figure*}[t!]
    \centering    \includegraphics[width=1\linewidth]{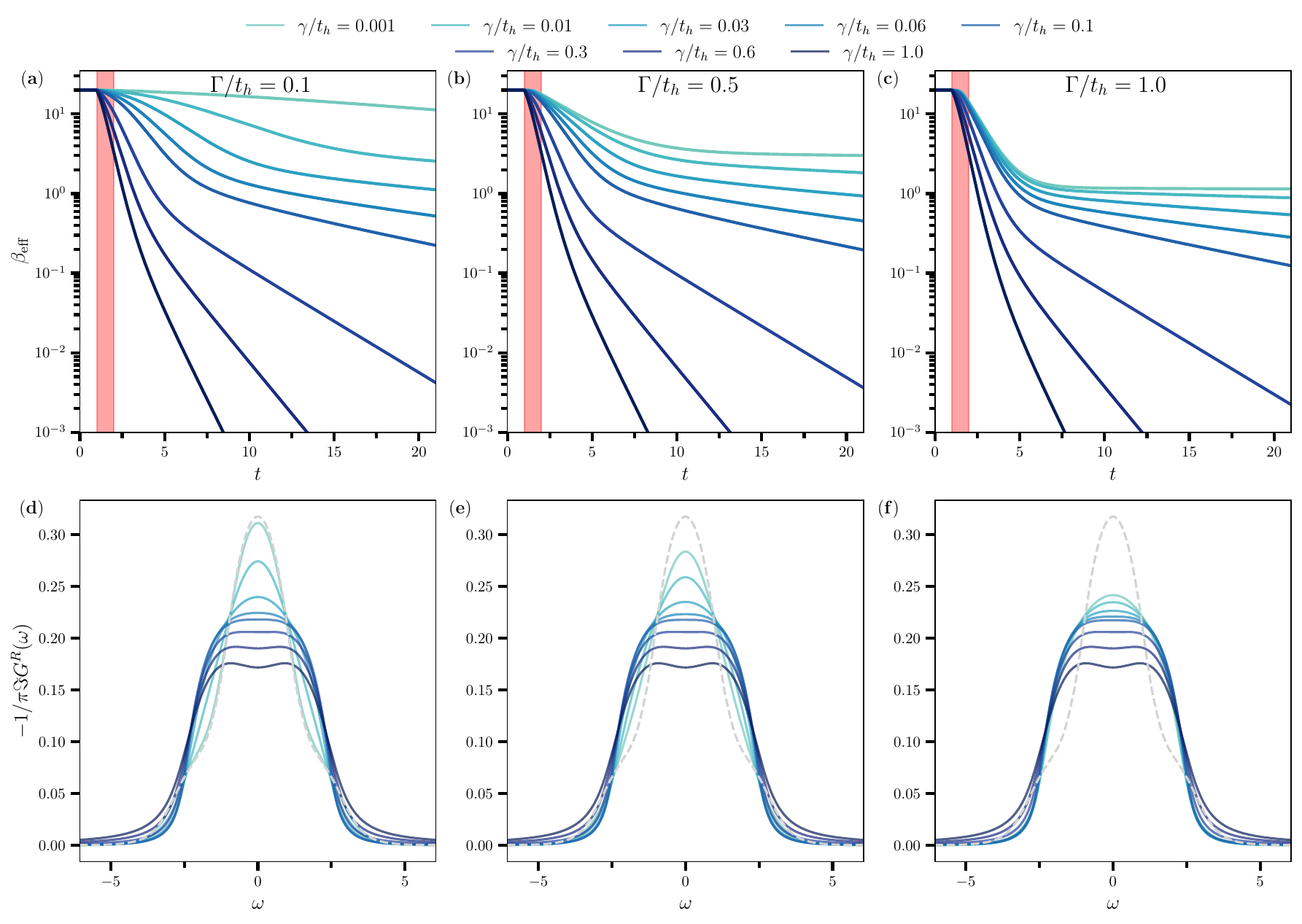}
    \caption{Dissipative Fermi-Hubbard Model under photo-excitation - (a-c) Dynamics of the effective temperature as a function of the dephasing rate $\gamma$ for increasing value of photo-excitation amplitude $\Gamma / t_h = 0.1,0.5,1.0$. (d-f) corresponding prethermal spectral function, where the gray dot represents the non-dissipative case. Here the Coulomb interacting is fixed to $U/t_h = 2$.
}
    \label{fig:QuenchPhotoexcitation}
\end{figure*}
%%%%%%%%%%%%%%%%%%%%%%%%%%%%%%%%%%%%%%%%%%%%%%%%%%%%%%%%%%%%%%%%%%%%

In this section we consider a different protocol, where the system is driven out of equilibrium via a photo-excitation and a simultaneous sudden switching of the dissipation $\gamma$. To simulate a photo-doping excitation, the system is transiently coupled to a fermionic bath with a suitably chosen spectral density. This in practice gives rise to an additional contribution to the electronic self-energy in terms of a kernel $\Gamma(t,t')$
\begin{align}
    \Gamma(t, t') = V(t) G_\text{bath}(t, t') V(t')^*,
\end{align}
where $V(t)$ is the time-dependent profile and $G_{\rm bath}$ the Green function of the fermionic bath (see Appendix~\ref{sec:ex_prot} for more details concerning the excitation protocol). At the same time, $t=0^+$, the dissipation $\gamma$ is suddenly switched from  $\gamma=0$ to $\gamma\neq0$. This double excitation or double quench is often used for nonequilibrium problems~\cite{mitra2011mode,schiro2014transient,larzul2022quenches}. It provides us with a knob to tune the degree in which the system is pushed out of equilibrium by a coherent or a dissipative perturbation, thus allowing to appreciate the interplay between the two processes.

%The time-dependent transfer of the electrons from the lower to the upper energy band is simulated by coupling an electron reservoir with occupied density of
%states $\mathcal{A}^<_{\rm bath}(\omega)$ at positive energies and unoccupied density of states $\mathcal{A}^>_{\rm bath}(\omega)$ at negative energies.  The coupling lasts for only one hopping time and happens at $t = 0^+$ (see App.~\ref{sec:ex_prot}
%for details concerning the excitation protocol).

In Fig.\ref{fig:QuenchPhotoexcitation}(a–c), we present the dynamics of the effective temperature for increasing values of the dissipation rate $\gamma$ and photo-excitation amplitude $\Gamma$. We observe that when the dissipation rate is on the order of the photo-excitation amplitude, the effective temperature dynamics remain largely unchanged, meaning that the time scale of the dissipation is faster than the photo-excitation, which is confirmed by looking at the prethermal spectral function in Fig.\ref{fig:QuenchPhotoexcitation}(d-f).
In contrast, when the dissipation rate is much smaller than the photo-excitation amplitude, $\gamma \ll \Gamma$, the dynamics become strongly dependent on $\Gamma$. Specifically, the dynamics accelerates as $\Gamma$ increases. More notably, the prethermal plateau becomes more stable and persists over a longer time window compared to the case without photo-excitation, as illustrated in Fig.~\ref{fig:Quench_gamma_Effective_temperature}(b). Overall, these results confirm that, largely independently from the type of non-equilibrium excitation, the dissipative Fermi-Hubbard model thermalizes within DMFT to an infinite temperature state. 
The details of the protocol matter for what concerns the thermalization pathways, the timescales required to reach the infinite temperature steady-state, and the possible existence and lifetime of prethermal metastable states. 
%and their lifetime. 
In the next section, we take a closer look at how thermalization under open-system dynamics arises in DMFT.

%In Fig.~\ref{fig:QuenchPhotoexcitation}(a-c), we show the dynamics of the effective temperature for increasing value of dissipation $\gamma$ and photo-excitation amplitude $A$. We that for dephasing of the order of the photoexcitation Amplitude the dynamics of the effective temperature is mostly unchanged, meaning that the time scale of the dissipation is faster than the photoexcitation, which is confirmed by looking at the prethermal spectral function in Fig.\ref{fig:QuenchPhotoexcitation}(d-f). However, for dephasing rate much smaller than the photo-excitation amplitude $\gamma \ll A$, we see that the dynamics strongly depends on A. The latter accelerate as we increase A and more interesting the  prethermal plateau becomes more stable and on a longer time-windows, compared to the case $A=0$ in Fig.~\ref{fig:Quench_gamma_Effective_temperature}(b). 

\section{Emergence of Infinite Temperature Thermalization}\label{sec:front}

In this section we discuss more in detail how the onset of infinite temperature thermalization emerges within DMFT. 
To this extent we take advantage of recent progress in understanding thermalization of closed quantum systems. In Ref.~\cite{Picano2025} we have shown that an interacting isolated quantum many-body system, such as the Fermi-Hubbard model, acts as \emph{its own thermal bath} making local observables reach thermal equilibrium after a nonequilibrium perturbation. This process emerges naturally by looking at DMFT equation from a different perspective, the one in which the impurity is self-consistently updated together with the bath. In practice, instead of making the DMFT converge at each time step, we first evolve the dynamics of the impurity in a given bath up to long times. Then, by imposing the self-consistency, we obtain the new bath for the next iteration $n$ (see Appendix~\ref{sec:StepByStep_QBE} for further details).
This amounts, in practice, in exchanging the order of limits between $t\rightarrow\infty$ (long-time limit) and $n\rightarrow\infty$ (DMFT self-consistency).
\begin{figure*}[t!]
    \centering    \includegraphics[width=1\linewidth]{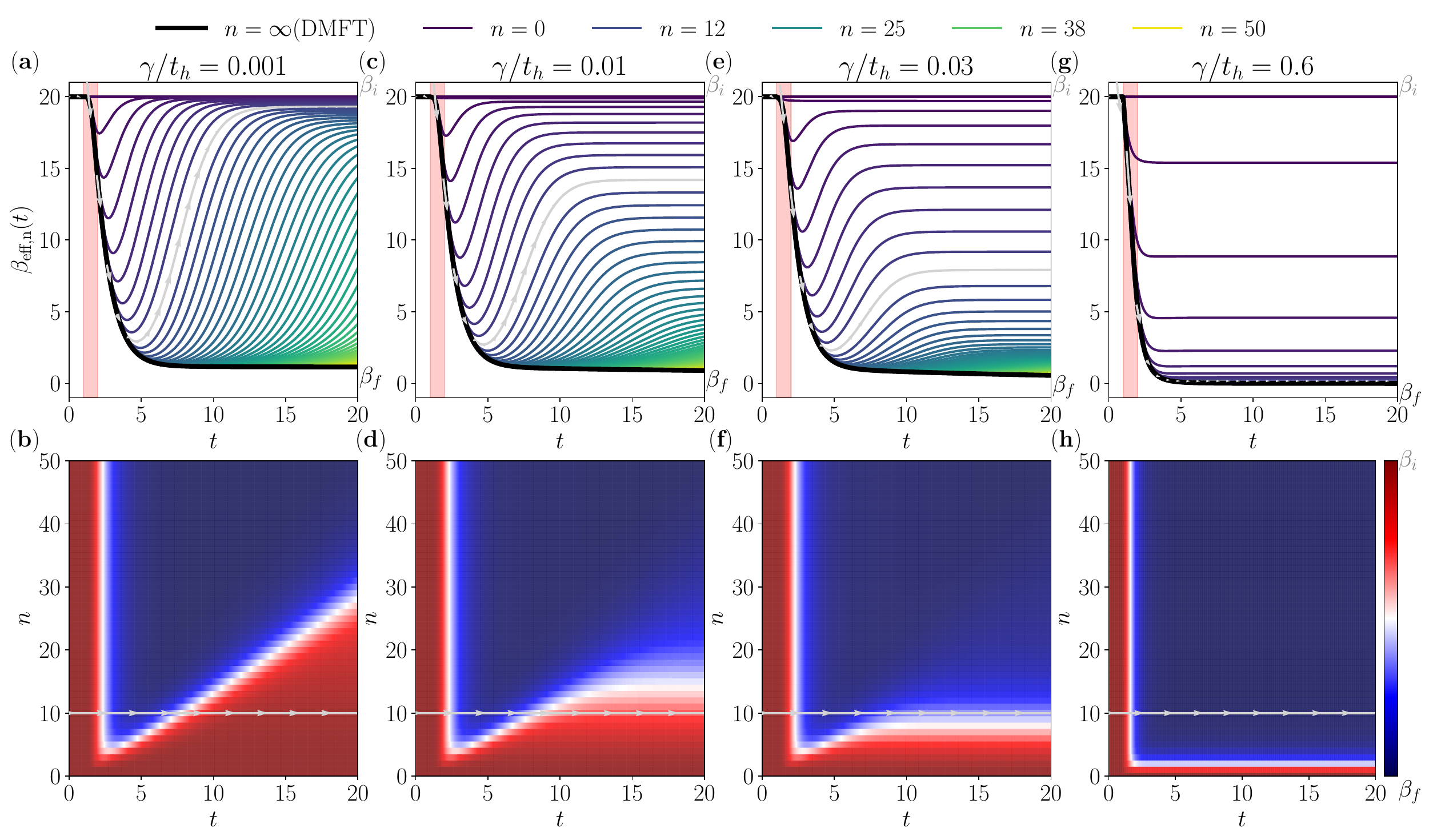}
    \caption{Thermalization Front of the Dissipative Hubbard Model for the Effective Temperature - Panel (a-h) describes the dynamics of the effective
    temperature for each iteration n and for different values of dephasing $\gamma/ t_h= 0.001,0.01,0.03,0.6$, compared to the fully self-consistent DMFT solution (black line) as obtained with QBE. We see clearly that contrary to the unitary Hubbard model, the effective temperature do not goes back to
    the initial value for all the iteration n, but reaches a non-thermal steady-state at $\beta_{\rm eff,n} < \beta_{i}$. This non-thermal steady-state act as a cut-off of the thermalization front. Here the photo-excitation amplitude is fixed at $\Gamma/ t_h =1$.
    %\textcolor{red}{Should we change a bit the colobar ? since here we have the impression that we have basically 2 state the initial and the fully converged - we do not see all the prethermal $\beta_n$ (as in the other plot of U=3)}
    }
    \label{fig:DMFTstepbystep}
\end{figure*}

This analysis, applied to an isolated system evolving unitarily, revealed the emergence of a sharp \emph{thermalization front} in the plane $(n,t)$ separating the initial state (where the memory of the initial condition is) from the full thermal state~\cite{Picano2025}. This has a simple interpretation: under a local perturbation the impurity at long-times always thermalizes back to the temperature of the bath, i.e. the initial temperature; still, the DMFT self-consistency heats up the bath more and more as $n$ increases, by propagating a thermal wave from short to long times. 

We now use the same approach for our dissipative Hubbard model with dephasing, namely we solve the DMFT equations step by step following the bath as it adapts to the impurity and viceversa, with the goal of understanding how the DMFT self-consistency allows the lattice model to heat up to infinite temperature.

In Fig.~\ref{fig:DMFTstepbystep}(a–g, top panels), we show the dynamics of the effective temperature $\beta_{\rm eff}$ at different DMFT iteration steps, and for increasing values of the dephasing rate $\gamma / t_h = 0.001,0.01,0.03,0.6$ (from left to right). These results are compared to the fully converged DMFT solution ($n=\infty$), shown with a dark full line. 

For very weak dissipation, $\gamma / t_h=0.001$, we see that the effective temperature decreases in time (following essentially the fully converged DMFT solution), then reaches a minimum and starts increasing, going back to a value which is very close to the initial one. If we plot these data in the $n,t$ plane (see bottom panels) we see clearly a sharp thermalization front with a linearly dispersing edge, characteristic of ballistic spreading, in accordance with the results of the unitary case~\cite{Picano2025}. As we increase the value of the dephasing however we see a qualitative change in this picture. Notably, we see that at long times the effective temperature $\beta_{\rm eff,n}(t)$ after each DMFT iteration does not reach the same initial value. Rather it stabilizes into a plateau value corresponding to a final temperature $\beta_f<\beta_{\text{eff,n}}<\beta_i$, where $\beta_i$ and $\beta_f$ are the initial and final inverse temperatures of the full-DMFT solution, shown in the top panels of Fig.~\ref{fig:DMFTstepbystep}. As the DMFT self-consistency is iterated this plateau value decreases with $n$, ultimately reaching $\beta_{\rm eff,n}\sim \beta_f$ for $n\rightarrow\infty$. We note that for non-interacting dissipative fermions the approach with $n$ is clearly exponential, while in presence of interaction one can see a richer behavior with metastable states (see Appendix \ref{sec:beta_n}). As a result, we see that the sharp thermalization front gets strongly affected by dissipation, first bending and flattening as $\gamma$ increases and ultimately completely disappearing for sufficiently large dephasing (see panel h). 

We can understand this phenomenology by thinking again at the physics of the (dissipative) quantum impurity model onto which the Fermi-Hubbard model with dephasing is mapped. Here the dephasing acts only on the impurity site, which is also coupled to the non-Markovian DMFT bath. As such, from the point of view of the impurity, dephasing is not sufficient to drive the system towards infinite temperature nor it is enough to allow the system to thermalize back to the temperature of the bath as it was the case for the unitary case. As a result the dissipative quantum impurity model sets itself into a nonequilibrium steady-state, resulting from the simultaneous coupling with two different environments, leading to a finite effective temperature. Upon increasing $n$, as the DMFT bath is made more and more self-consistent by adjusting the bath spectrum to the local impurity Green's function, the impurity is heat up more and more and the next step effective temperature increases up to the point in which $\beta_{\rm eff,n}\rightarrow 0$. Increasing the value of the dephasing makes the process of thermalization to a nonequilibrium steady state more efficient and fast and makes the memory of the initial condition fade away more rapidly, with the result that the front bends more and more and ultimately disappears. We conclude therefore that for dissipative system thermalization does not arise as a thermal front, characterised by a memory of the initial condition, but rather as a single step process. This is reasonable since the Lindblad dynamics involves irriversibility and does not retain memory of the initial condition. Our results therefore confirm that this approach captures some key feature of the nonequilibrium dynamics.

%we see a clear thermalization front

%After a sharp decrease at short times, corresponding to the interval of the photo-excitation and the quench of the dephasing, the dynamics of $\beta_{\rm eff}$ decays, reaching a plateau value (thick black line) corresponding to the fully converged DMFT solution. 

%At finite DMFT iterations n, the dynamics of the effective temperature $\beta_{\rm eff}$ initially follows that of the fully converged solution but eventually deviates at longer times to reach a non-equilibrium prethermal plateau. This non-monotonic behavior of the effective temperature arises from the competition between energy injection by the photoexcitation and energy loss due to Markovian dissipation. During the photo-excitation process, energy is rapidly pumped into the system, causing the effective temperature to rise. Then after the pulse ends, the dissipation dominates and gradually tend to reach a dark state of the quantum impurity model.

%\textcolor{red}{Need to be finish.Discuss how $\beta_{n}$ depend on $\gamma$ ?}

We close by noting  that this thermalization front and its lack of at strong dephasing is a direct consequence of 
having a perturbation acting on the unitary part of the evolution, in our case the photo-excitation. Indeed when only the dephasing is quenched, the non-reversible nature of the dissipation leads to a monotonic decrease in the effective temperature over time, eventually reaching a prethermal plateau at $\beta_{\rm{eff},n}$. We expect that performing a quench in the interaction, instead of photoexcitation, would lead to qualitatively similar results.

\section{Conclusions}
\label{sec:conclusion}

In this work we have studied a dissipative Fermi-Hubbard model with dephasing and discuss the dynamics of thermalization and heating due to dissipation. We have solved the many-body Lindblad dynamics using nonequilibrium DMFT, corresponding to take the infinite connectivity limit of the lattice problem and solving a self-consistent dissipative quantum impurity model. The latter in the present case takes the form of a dissipative Anderson impurity model with dephasing, which we have solved with weak-coupling perturbation theory.  We have focused on two different nonequilibrium protocols, corresponding to (i) a pure quench of the dissipation and (ii) a simultaneous photo-excitation and dissipation quench. 

We have discussed the heating dynamics of the system, as described by the kinetic and potential energy as well as the effective temperature, which show thermalization towards an infinite temperature state. We have shown that the heating rates depend strongly on interaction and dephasing. In particular, for the case of dissipation quench we have shown the emergence of a prethermal regime for weak dissipation, where a quasi-stationary state is formed before heating dominates at long times. This prethermal steady state is shown to still display coherent quasiparticles and a quasi-equilibrium distribution function at low energy, while high-frequency excitations are non-thermally occupied. The dynamics under protocol (ii) corresponding to simultaneous photo-excitation and dephasing display similar features, with the addition that the strength of the photodoping has the effect of stabilizing a prethermal plateau for weak dissipation.

Finally we have studied how thermalization to infinite temperature arises from the point of view of the impurity, or equivalently how the many-body system acts as its own thermal bath, using recent insights obtained from the closed system case. We have shown that the thermalization front emerging in the unitary case is strongly modified by dissipation, a signature of the irreversibility of the open system dynamics. In particular, while for weak dephasing a linear front remains visible, for moderate and large values of dissipation the front bends and flatten out, a signature of the fact that the thermalization dynamics becomes completely dominated by the dissipative processes. 

In the future it would be interesting to explore the effect of different types of dephasing mechanism, as done for the impurity model in Refs.~\cite{vanhoecke2024diagrammatic,vanhoecke2025kondozeno}, as well as the regime of strong interaction and strong dissipation, where the perturbative approach used in this work would not work.

\begin{acknowledgments}
We thank Denis Golež for valuable and stimulating discussions.
We acknowledge financial support
from the ERC consolidator Grant No. 101002955 - CONQUER. A. P. acknowledges funding from the
European Union’s Horizon 2020 research and innovation programme under the Marie Sklodowska-Curie Postdoctoral
Fellowship (Grant
Agreement No. 101149691 - DISRUPT). We are grateful for the use of computational resources from the Collège de France IPH cluster. A. P. is grateful to Philipp Hansmann and the RRZE of the University of Erlangen-Nuremberg for providing additional computational resources.
%We acknowledge the computational resources on the Coll\`{e}ge de France IPH cluster.
\end{acknowledgments}

\appendix

\section{From Lindblad to Out-of-Equilibrium DMFT}
\label{sec:Appendix_Lindblad_to_DMFT}
We present here the formalism of DMFT for Markovian dissipation.
As already explained in the main text, the starting point to derive non-equilibrium DMFT equations for Markovian systems is to cast the Lindblad master equation~\eqref{eqn:Lindblad} in the language of non-equilibrium Keldysh field theory, given the Hubbard Hamiltonian \eqref{eqn:Hubbard} and the local jump operators in Eq.~\eqref{eq:jump_dephasing}.
In this formalism, the evolution of the system is encoded in the partition function $\mathcal{Z}$ in Eq.~\eqref{eqn:Z}, with the Keldysh action \eqref{eqn:S_imp}. The coherent and dissipative contributions to the dynamics are captured through the Lindbladian term $\mathcal{L}$ in Eq.\eqref{eqn:Lindblad_Keldysh}.
% The Keldysh partition function of the dissipative Hubbard model is given by 
% \begin{align}
%     \mathcal{Z} = \int \prod_{i,\sigma} \mathcal{D}\left[ c_{i\sigma},\bar{c}_{i\sigma}\right] e^{iS}
% \end{align}
% where $\mathcal{S}$ is the Keldysh action
% \begin{align}
% 	\label{eqn:S_imp}
%     \mathcal{S} = \int_{-\infty}^{+\infty} dt \sum_{\alpha=\{-,+\}} \sum_{i\sigma} \alpha \bar{c}_{i\sigma \alpha} i\partial_t c_{i\sigma \alpha} -i\int_{-\infty}^{+\infty} dt \mathcal{L}
% \end{align}
% defined on the Keldysh contour $\{+,-\}$ in Fig.~\ref{fig:Keldysh}.
% In the previous equation, we used the Grassmann variables $\bar{c}_{i\sigma \alpha}$ and $c_{i\sigma \alpha}$. There, $\alpha$ corresponds to the Keldysh index,  $\alpha \in \{+,-\}$; $\sigma$ represents the spin index, and $i$ to the site index. 
% The Lindbladian in Eq.~\eqref{eqn:S_imp}, in Keldysh formalism, reads:
% \begin{align}
%     \mathcal{L} = -i \left( H_{ +}-H_{ -}\right) + \sum_{i,\mu} L_{i\mu,+}  \bar{L}_{i\mu,-} - \frac{1}{2} \left( \bar{L}_{i\mu,+}L_{i\mu,+}+\bar{L}_{i\mu,-} L_{i\mu,-}\right)
% \end{align}
We split the action $\mathcal{S}$ in Eq.~\eqref{eqn:S_imp} into three parts,
\begin{align}
    \mathcal{S} = \mathcal{S}_{ \rm loc} + \delta \mathcal{S} + \m{S}_{(0)}
\end{align}
where the first contribution $\mathcal{S}_{\rm loc}$ corresponds to the part of the action, that only contains grassmann at site $i=0$,
\begin{align}
    S_{\rm loc} &= \int dt \sum_{\alpha \sigma}  \alpha  \left( \bar{c}_{0\sigma \alpha} i\partial_t c_{0\sigma \alpha} -  U  n_{0\uparrow\alpha} n_{0\downarrow\alpha} \right) \nonumber \\
    &-i \left(\sum_{\sigma} L_{0\sigma+}  \bar{L}_{0\sigma-} - \frac{1}{2} \left( \bar{L}_{0\sigma+}L_{0\sigma+}+\bar{L}_{0\sigma-} L_{0\sigma-}\right) \right)
\end{align}
Then the second term $\delta \m S$ takes into account the hopping between the site $i=0$ and the other sites of the lattice, $i\neq 0$,
\begin{align}
    \delta \m S = -\frac{t_h}{\sqrt{z}} \int dt \sum_{\alpha \sigma} \sum_{\langle 0, j \rangle} \alpha \left( \bar{c}_{0\sigma \alpha} c_{j\sigma \alpha} + \bar{c}_{j\sigma \alpha} c_{0\sigma \alpha}  \right)
\end{align}
Finally the last contribution $\m{S}_{(0)}$ is the part of the action where the site $i=0$ and
its bonds are removed  ($i,j \neq 0$)
\begin{align}
    S_{(0)} &= \int dt \sum_{\alpha \sigma} \sum_{i\neq 0} \alpha  \left[ \bar{c}_{i\sigma \alpha} i \partial_t c_{i\sigma \alpha}  - \alpha H_{(0)\alpha}\right] \nonumber \\
    &-i \sum_{i\neq 0, \sigma} L_{i\sigma+}  \bar{L}_{i\sigma-} - \frac{1}{2} \left( \bar{L}_{i\sigma+}L_{i\sigma+}+\bar{L}_{i\sigma-} L_{i\sigma-}\right)
\end{align}
with $H_{(0)}$ the Hamiltonian defined as,
\begin{align}
    H_{(0)} = - \frac{t_h}{\sqrt{z}} \sum_{\langle i,j \rangle, i,j\neq 0 ,\sigma}  \left( c_{i\sigma}^\dagger c_{j\sigma} + c_{j\sigma}^\dagger c_{i\sigma}  \right) + \sum_{i\neq 0} U n_{i,\uparrow} n_{i\downarrow}
\end{align}
The first step of the derivation is to rewrite the partition function $\m{Z}$ in terms of the three contributions of the Keldysh action,
\begin{align}
    \mathcal{Z} &= \int \prod_{i ,\sigma} \mathcal{D}\left[ c_{i\sigma},\bar{c}_{i\sigma}\right] e^{i \m S} \nonumber \\ 
    &= \int \prod_{\sigma} \mathcal{D}\left[ c_{0\sigma},\bar{c}_{0\sigma}\right] e^{i \m S_{\rm loc}} \int \prod_{i\neq 0,\sigma} \mathcal{D}\left[c_{i\sigma},\bar{c}_{i\sigma}\right] e^{i \m S_{(0)}} e^{i \delta \m S}
\end{align}
We now introduce the contour-ordered expectation value with respect to the action $S_{(0)}$,
\begin{align}
    \langle X \rangle_{(0)} = \frac{1}{\mathcal{Z}_{(0)}} \int \prod_{i\neq 0, \sigma} \mathcal{D}\left[ c_{i\sigma},\bar{c}_{i\sigma}\right] X e^{i \m S_{(0)}}
\end{align}
where $\mathcal{Z}_{(0)}$ is the corresponding partition function of the decoupled lattice. Using this relation, the full partition function can be recast as
\begin{align}
    \m{Z} = \m{Z}_{(0)} \int \prod_\sigma \m{D}\left[ c_{0\sigma},\bar{c}_{0\sigma}\right] e^{i \m S_{\rm loc}} \langle  e^{i\delta \m S}\rangle_{(0)}
\end{align}
By using a cumulant expansion, we can formally rewrite the expectation value $\langle e^{i \delta \m S}\rangle_{(0)}$ as,
\begin{align}
    \langle e^{i \delta \m S}\rangle_{(0)} &= \exp \big [ \langle i\delta \m S \rangle_{(0)} \nonumber \\ 
    &+ \frac{1}{2} \left( \langle \left[ i\delta \m S \right]^2 \rangle_{(0)} - \left[\langle i\delta \m S \rangle_{(0)}\right]^2 \right) +\cdots \big ]
\end{align}
with,
\begin{align}
    \langle (i \delta \m S)^2 \rangle_{(0)} &=  2\int d \tau \int d \tau^\prime \sum_{\alpha \beta} \alpha \beta   \nonumber \\
    &\sum_{\sigma }   \bar{c}_{0\sigma \alpha} (t_1)\left[\sum_{j,k \neq 0} \frac{t_h^2}{z}  G_{jk, \sigma}^{\alpha \beta,(0)} (t_1,t_2)\right] c_{0\sigma \beta } (t_2)
\end{align}
where we defined the Green's function $G_{jk, \sigma}^{\alpha \beta (0)}$ as,
\begin{align}
    G_{jk, \sigma}^{\alpha \beta (0)} (t,t^\prime) = - i \langle \m{T}_\m{C} c_{j\sigma \alpha}(t) \bar{c}_{k\sigma \beta}(t^\prime)\rangle_{(0)}
\end{align}
with $T_{\m{C}}$ the time-ordering operator along the Keldysh contour. 

The central idea of dynamical mean-field theory (DMFT) is to derive an effective Keldysh action for the fermionic degrees of freedom on a single lattice site (the "impurity") by integrating out all other sites in the system, thus mapping the fully interacting lattice model onto a single interacting impurity coupled to a self-consistently determined quantum bath. In the large connectivity limit, $z\gg 1$, one can expand the action in power of $1/z$, obtain the effective Keldysh action, and write it in closed form as,
\begin{align}
    \label{Seff}
    & \m S_{\rm eff} \left[\bar{c}_{0\sigma \alpha} , c_{0\sigma \alpha}  \right] = \log \int \prod_{i \neq 0} \m{D}\left[\bar{c}_{i\sigma},c_{i\sigma} \right] e^{i \m S} \notag \\ &=  \m S_{\rm loc} - i \sum_{\sigma} \sum_{\alpha \beta}\alpha \beta  \int dt dt^\prime \bar{c}_{0\sigma \alpha}(t) \Delta_\sigma^{\alpha \beta}(t,t^\prime) c_{0\sigma\beta}(t^\prime)
\end{align}
In this expression, $\m S_{\rm loc}$ is the local on-site contribution of the original lattice action Eq.~\eqref{eqn:S_imp}, which includes interactions and possible dissipation. The second term captures the influence of the rest of the lattice on the impurity site via an effective non-Markovian bath characterized by the hybridization function ,$\Delta_\sigma^{\alpha \beta}$  which is given by:
\begin{align}
    \Delta_\sigma^{\alpha \beta} (t,t^\prime) = \sum_{j,k\neq 0} \frac{t_h^2}{z} G_{jk, \sigma}^{(0),\alpha \beta} (t,t^\prime)
\end{align}
To close the DMFT loop, one must impose a self-consistency condition that relates observables computed from the cavity action to those obtained from the effective impurity action \eqref{Seff}. The specific form of this condition depends on the lattice geometry. For the Bethe lattice this relation becomes especially simple, as the removal of a site fully decouples its neighbors. In this case, the off-diagonal cavity Green's functions vanish, and we obtain:
\begin{align}
    \label{eqn:bethe_selfcons}
    G_{jk, \sigma}^{(0)} (t,t^\prime) = \delta_{j,k} G_{jk, \sigma}^{S_{\rm eff}} (t,t^\prime)
\end{align}
Using this, the hybridization function simplifies to
\begin{align}
	\label{Delta}
\Delta_\sigma^{\alpha \beta} (t,t^\prime) = \frac{t_h^2}{z} G_{ \sigma}^{S_{\rm eff}, \alpha \beta} (t,t^\prime)
\end{align}
which directly relates the hybridization function of the non-Markovian bath to the interacting Green's function of the impurity,
\begin{align}
	\label{Gimp}
	G^{S_{\rm eff, \alpha \beta}}_\sigma (t,t^\prime) = -i \langle c_{0\sigma \alpha} (t) \bar{c}_{0\sigma \beta} (t^\prime)  \rangle_{S_{\rm eff}} 
\end{align}
Solving the original dissipative and interacting lattice problem within DMFT thus amounts to self-consistently solving the impurity problem defined by the effective Keldysh action \eqref{Seff}, computing the corresponding impurity Green's function \eqref{Gimp}, and updating the hybridization function via Eq.~\eqref{Delta} until convergence is achieved.

\section{Derivation of the Dissipative Self-Energy } \label{sec:AppendixSelfEnergyDerivation}

In the previous section, we demonstrated that in the limit $z\rightarrow \infty$, the fully interacting and dissipative Hubbard model can be exactly mapped onto a dissipative quantum impurity model. Building on this result, we now focus on the dissipative part of self-energy for the non-interacting impurity solver. As a starting point, we consider the partition function of the impurity model, which takes the form,
\begin{align}
    \mathcal{Z} = \int \mathcal{D} [c, \bar{c}] \exp{ \{i \mathcal{S}_{\rm free}+i\mathcal{S}_{ \gamma} \} }
\end{align}
where $\mathcal{S}_{\rm free}$ denotes the action of the non-dissipative impurity model, and $\mathcal{S}_{ \gamma}$ represents the dissipative contribution, which we recall is given by:
\begin{align}
    S_{\gamma } = -i \left[ \int d \tau \sum_\sigma \gamma n_{\sigma +} n_{\sigma -} - \frac{\gamma}{2} n_{\sigma+}  - \frac{\gamma}{2} n_{\sigma-}  \right]
\end{align}
where $n_{\sigma,\pm} = c_{\sigma,\pm}^\dagger c_{\sigma,\pm} $ are the number operators of the impurity on the forward $(+)$ and backward $(-)$ branches of the Keldysh contour, respectively. From this partition function, the interacting impurity Green's function is defined as the contour-ordered expectation value
\begin{align}
    G_{\sigma}^{\alpha \beta}(t,t') &= -i \langle \mathcal{T}_\mathcal{C} c_{\sigma\alpha}(t) c_{\sigma\beta}^{\dagger}(t') \rangle_{\mathcal{S}= \mathcal{S}_{\rm free}+\mathcal{S}_{\gamma}} \nonumber \\
    &= -i \int \mathcal{D} [\psi, \bar{\psi}] c_{\sigma\alpha}(t) c_{\sigma\beta}^{\dagger}(t') e^{i \mathcal{S}_{\rm free}+i\mathcal{S}_{\gamma}} 
\end{align}
By using a standard Taylor expansion for the dissipative action $S_{\gamma}$, we can write all the order of the Green's function,
\begin{align}
    G_{\sigma}^{\alpha \beta}(t,t') = -i \int \mathcal{D} [\psi, \psi^\ast] e^{i \mathcal{S}_{\rm free}} 
    c_{\sigma \alpha}(t) c_{\sigma \beta}^{\dagger}(t') [1+i \mathcal{S}_{\gamma}+\cdots] 
\end{align}
which serves as a starting point for a perturbative treatment of the dissipation within the impurity problem.

\subsection{Perturbative expansion in terms of $\gamma$ }

At first order in the dissipation strength $\gamma$, , the correction to the Green's function can be decomposed into two distinct contributions,
\begin{align}
    G^{(1)\alpha \beta}_{\sigma} (t,t^\prime) = G^{(1)\alpha \beta}_{\sigma,\rm jump} (t,t^\prime) + G^{(1)\alpha \beta}_{\sigma,\rm NH} (t,t^\prime)
\end{align}
where $G^{(1)\alpha \beta}_{\sigma,\rm jump}$ arises from the quantum jump terms, and $G^{(1)\alpha \beta}_{\sigma,\rm NH}$ originates from the non-Hermitian part of the dissipative action. These contributions are explicitly given by,
\begin{align}
G^{(1)\alpha \beta}_{\sigma,\mathrm{jump}}(t,t') = \gamma \int d\tau \, \Big[ -  
& \mathcal{G}_\sigma^{\alpha +}(t, \tau) \, \mathcal{G}_\sigma^{--}(\tau, \tau) \, \mathcal{G}_\sigma^{+ \beta}(\tau, t') \notag \\
- & \mathcal{G}_\sigma^{\alpha -}(t, \tau) \, \mathcal{G}_\sigma^{++}(\tau, \tau) \, \mathcal{G}_\sigma^{- \beta}(\tau, t') \notag \\
+ & \mathcal{G}_\sigma^{\alpha +}(t, \tau) \, \mathcal{G}_\sigma^{+-}(\tau, \tau) \, \mathcal{G}_\sigma^{- \beta}(\tau, t') \notag \\
+ & \mathcal{G}_\sigma^{\alpha -}(t, \tau) \, \mathcal{G}_\sigma^{-+}(\tau, \tau) \, \mathcal{G}_\sigma^{+ \beta}(\tau, t') 
\Big]
\end{align}
and
\begin{align}
    G^{(1)\alpha \beta}_{\sigma,\rm NH} (t,t^\prime) = \gamma \sum_{\alpha_1 = +,-} \int d \tau  \mathcal{G}_\sigma^{\alpha \alpha_1}  (t,\tau )\mathcal{G}_\sigma^{\alpha_1 \beta} (\tau,t^\prime)
\end{align}
where we have used the Wick’s theorem, since the non-dissipative part of the action $S_{\rm free}$ is quadratic.  By summing all first-order contributions, we obtain the full correction $G^{(1)\alpha \beta}_{\sigma}$ including the $(++)$ component as well as all other Keldysh components. From this result, the first-order correction to the self-energy can be extracted via direct comparison with the Dyson equation,
\begin{align}
    \Sigma_{\gamma \sigma}^{(1)}(t,t^\prime) = \gamma \m{G}_{\sigma}(t,t) \delta(t-t^\prime)
\end{align}
In the same spirit, the second-order correction to the self-energy can be computed using higher-order terms in the perturbative expansion. Remarkably, for the specific choice of local dephasing considered here, it has been shown that the entire perturbative series can be resummed. This resummation yields an exact expression for the dissipative part of the self-energy, which takes a closed form~\cite{ji.fe.22,vanhoecke2024diagrammatic}, 
\begin{align}
    \Sigma_{\gamma \sigma} (t,t^\prime) = \gamma G_\sigma (t,t) \delta(t-t^\prime)
\end{align}

\section{Non-interacting Fermi-Hubbard Model}
\label{sec:AppendixNoninteractingCase}
Now that we have derived the dissipative part of the self-energy, we can analyze the exact dynamics of the non-interacting ($U=0$) dissipative Hubbard model. Even in the absence of Coulomb interactions, the system remains non-Gaussian due to the structure of the dissipation, which can lead to nontrivial dynamics and non-equilibrium  steady state.
We start by considering the Dyson equation defined in Eq.~(\ref{Gfull})
\begin{align}\label{eqn:GreenFCT}
	G_{\sigma}^{-1}(\omega)= \omega - \Delta_\sigma(\omega) -\Sigma_{\text{int},\sigma}(\omega) 
\end{align}
where in this case the interacting self-energy is only given by the dissipative part $\Sigma_{\text{int},\sigma}(\omega) = \Sigma_{\gamma \sigma}(\omega)$ defined in Eq.~\eqref{eqn:KBE_SelfEnergy}. As discussed in the main text, we are considering the Bethe lattice with $t_h=1$, such that the hybridization function Satisfies
\begin{align}\label{eqn:AppendixBethe}
    \Delta_\sigma(t,t^\prime) = t_h^2 G_\sigma(t,t^\prime)
\end{align}
Inserting \eqref{eqn:AppendixBethe} in \eqref{eqn:GreenFCT} yields
\begin{align}\label{eqn:DMFT_equation_noninteracting}
    G_\sigma(\omega) = \frac{1}{\omega - \epsilon_d  - t_h^2G_\sigma(z) -\Sigma_{\gamma \sigma} (z)}
\end{align}
which is the DMFT equation for the non-interacting but dissipative Hubbard model.

\subsection{Analytical expression of the Retarded Green's Function}
From Eq.~(\ref{eqn:DMFT_equation_noninteracting}), the simple structure of the dissipative self-energy $\Sigma_{\gamma \sigma}(\omega)$, allows us to derive an exact analytical expression for the retarded Green's function. Specifically, for the retarded component, the DMFT equation takes the form:
\begin{align}
		\label{GR_gamma_anl}
		G^R_\sigma(\omega) = \frac{1}{\omega + i\eta   - t_h^2G^R_\sigma(\omega) +i \frac{\gamma}{2}}
\end{align}
where we have introduced $\eta \rightarrow 0^+$ and we have used the fact that $\Sigma_{\gamma \sigma}(\omega) = -i \gamma /2$. The resulting expression can be formally rewritten in polynomial form:
\begin{align}
	\left[ G^R_\sigma(\omega) \right]^2 - \frac{z  +i \frac{\gamma}{2}}{t_h^2} G^R_\sigma(\omega) + \frac{1}{t_h^2}=0
\end{align}
whose solutions can be obtained analytically as,
\begin{align}
	\label{GR_gamma}
		G^R(\omega) = \frac{1}{t_h^2}  \left[ \left( \omega + i\eta +i \frac{\gamma}{2} \right) \pm \sqrt{\left( \omega + i\eta  +i \frac{\gamma}{2}\right)^2 - 4 t_h^2 }   \right] 
\end{align}
In the same spirit, we can derive an analytical expression for the Keldysh component of the Green's function, at least in the half-filling case, since dephasing does not break particle-hole symmetry.

\subsection{Steady-State Spectrum and Occupation}

\begin{figure}
    \centering
    \includegraphics[width=1\linewidth]{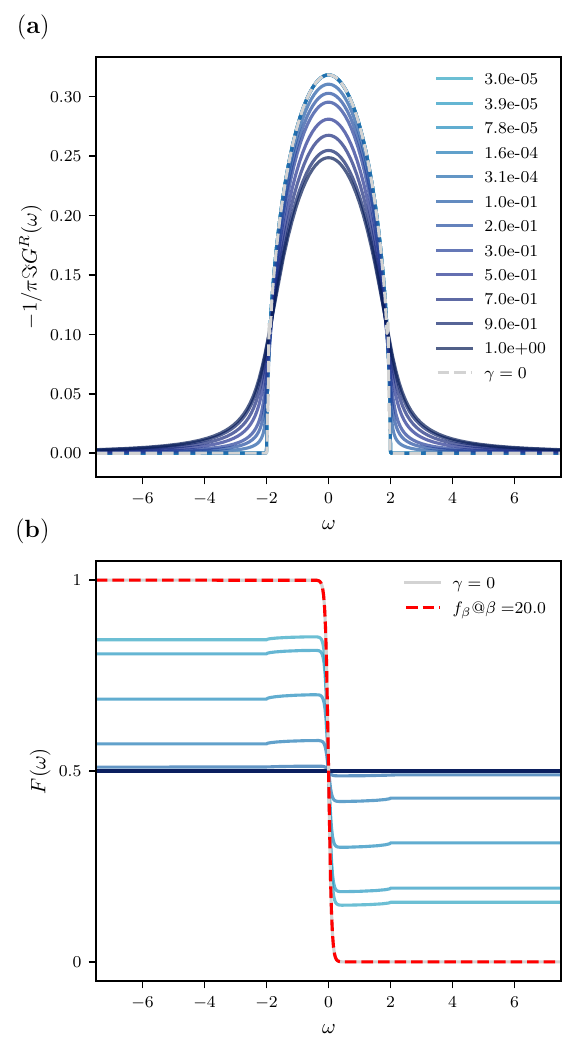}
    \caption{
    Dissipative Fermi-Hubbard Model - Spectral properties and distribution function of the non-interacting Fermi-Hubbard model ($U/t_h= 0$) for increasing values of dephasing $\gamma$ at $t=12500$. Panel (a) Spectral function in the prethermal steady-state as a function of the dissipative rate $\gamma$. Panel (b) shows the corresponding distribution function, with the dot red line the equilibrium fermi function at $\beta_i=20$.}
\label{fig:spectrum_distribution_full_U_0}
\end{figure}

We now turn our attention on the spectral function and the distribution function of the non-interacting dissipative Hubbard model, that we show in Fig.~\ref{fig:spectrum_distribution_full_U_0}, for increasing values of dephasing rate $\gamma$.  
As $\gamma$ increases, we see that the typical resonance of the bethe lattice is broadened and at the same time the value at $\omega=0$ decreases with $\gamma$, which is a signature of loss of coherence due to the Markovian environment. At the same time, the distribution function deviates from the equilibrium Fermi-Dirac form, reflecting the fact that the system is driven out of equilibrium and relaxes to a prethermal state. Importantly, it's seem that the dissipation act almost uniformly across all frequencies, meaning that the dissipation do not discriminate between low and high-energy excitations, it affects all many-body states with equal strength, regardless of their energy.  This energy-insensitive behavior is typical of dephasing processes, which tends to push the system toward a fully mixed state, similar to an infinite-temperature in the infinite time limit.

%\subsection{Step-by-step DMFT}

\section{Step-by-step Quantum Boltzmann Equation}
\label{sec:StepByStep_QBE}

In the QBE described in Sec.~\ref{sec:QBE}, given the distribution function $F(\omega,t)$, the spectrum $\mathcal{A}$, the hybridization function $\Delta$ and the self-energy $\Sigma$ are updated until  DMFT self-consistency is reached. 
Only after convergence, the distribution function  $F(\omega,t+h)$ is calculated with the QBE~\eqref{QBE}, and the new NESS loop starts. 
\begin{figure*}[t!]
    \centering
\includegraphics[width=0.8\linewidth]{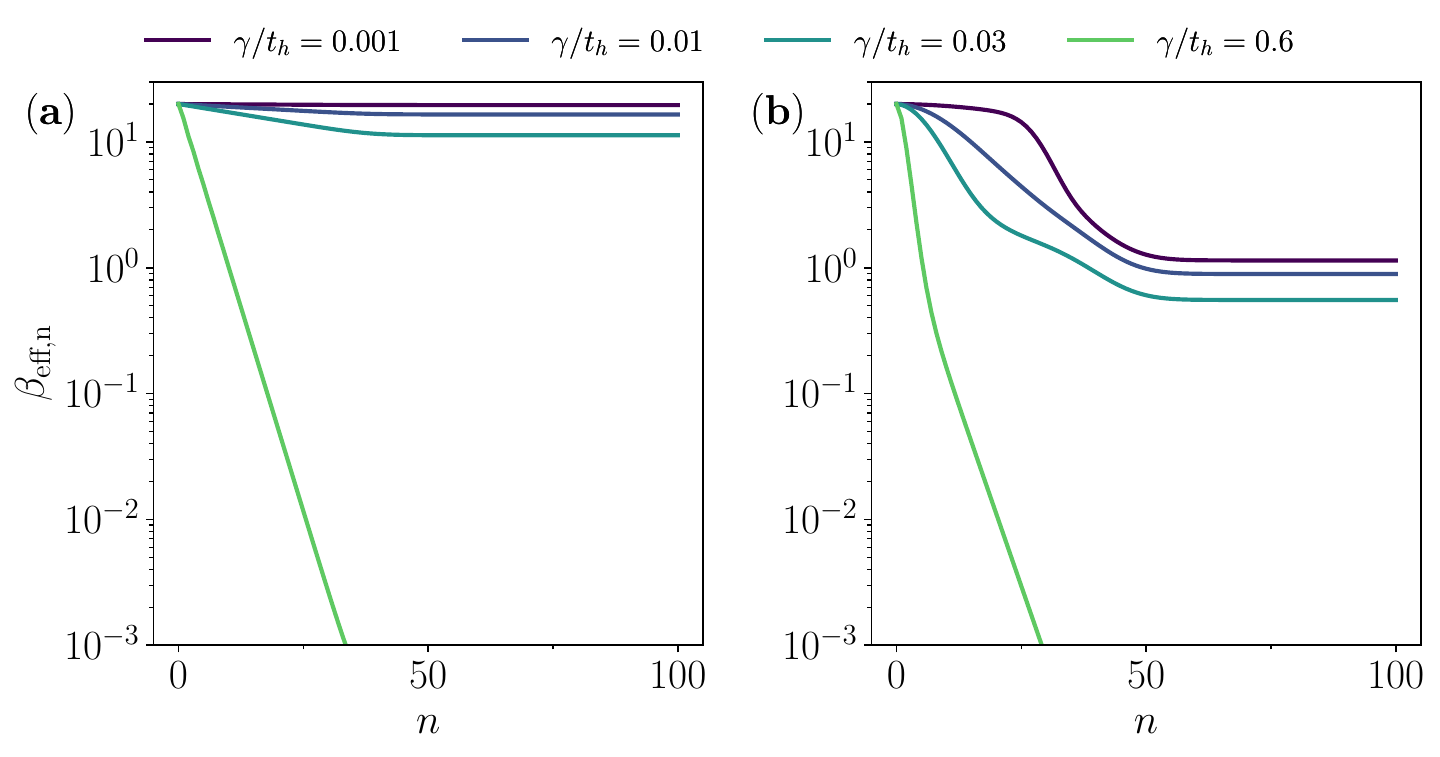}
    \caption{Emergence of Infinite Temperature Thermalization - 
    Evolution of the steady-state effective temperature in the step-by-step DMFT approach, solved with QBE, as a function of the DMFT iteration $n$ for different values of dephasing $\gamma/ t_h= 0.001,0.01,0.03,0.6$. The Hubbard interaction is $U/ t_h=0$ in panel (a) and $U/ t_h=2$ in panel (b).
    In both panels, the photo-excitation amplitude is fixed at $\Gamma/ t_h =1$ and the effective temperature is taken at $t=20$.   
    }
    \label{fig:prethermal_steadystate_beta}
\end{figure*}
Here, we consider the case in which DMFT convergence is reached globally in time rather than for each single timestep. We explain how it works.
At the first global DMFT iteration, $n=0$, the system is in equilibrium at the initial temperature $\beta_i$.
The equilibrium functions $G_{n=0}(\omega,t=0)$, $\Delta_{n=0}(\omega,t=0)$,  and $\Sigma_{n=0}(\omega,t=0)$ are determined from the same NESS loop described in the  Sec.~\ref{sec:QBE} in step 1) - 5).
%, and are called $G_{0}(\omega)$, $\Delta_{0}(\omega)$,  ,and $\Sigma_{0}(\omega)$.  
At the first global DMFT iteration ($n=0$), the excitation bath $\Gamma$ is set to zero for all the timesteps, and the equilibrium Green's functions are not updated in time: 
\begin{equation}
\begin{cases}
G_{n=0}(\omega,t) \equiv G_0(\omega) ,\,\,\,\, \forall t \nonumber \\
\Delta_{n=0}(\omega,t)  \equiv \Delta_0(\omega) ,\,\,\,\, \forall t \nonumber \\
\Sigma_{n=0}(\omega,t)  \equiv  \Sigma_0(\omega) ,\,\,\,\, \forall t
\end{cases}
\end{equation}
Starting from these functions, several global DMFT iterations $n$, each of which goes from $t=0$ to $t=t_{\text{max}}$, are performed until global DMFT convergence is reached. Each full time propagation $t \in [0,t_{\text{max}}]$ is identified by the index $n$ ($n \geq 1$). 
They all start from the same initial condition at $t=0$:
\begin{equation}
	\begin{cases}
	G_n(\omega,t=0)=G_0(\omega) ,\,\,\,\, \forall n \geq 1 \nonumber \\
	\Delta_n(\omega,t=0)=\Delta_0(\omega) ,\,\,\,\, \forall n \geq 1 \nonumber \\
	\Sigma_n(\omega,t=0)=\Sigma_0(\omega) ,\,\,\,\, \forall n \geq 1
	\end{cases}
\end{equation}
Given this initial condition, the DMFT loop for $t \in [0,t_{\text{max}}]$, for each full time propagation $n \geq 1$, looks like:
\begin{enumerate}
	\label{NESSloop1}
	\item 
	Update the retarded Green's function:
	\begin{align}
	 G_n^R(\omega,t)
	= 
	&[\omega +i0^+ -H_{\text{loc},n}(t) - \Delta_n^R(\omega,t) \nonumber \\ &-\Sigma_{\text{int},n}^R(\omega,t)]^{-1} ,\,\,\,\, \forall n \geq 1
	\end{align}
	and determine $  \mathcal{A}_n(\omega,t)=-\frac{1}{\pi} \Im  G^R_n(\omega+i0,t)$.
	We recall that in our case (half-filling) it is always $H_{\text{loc},n}(t) =0$. 
	\item
	Determine the lesser Green's function from the given distribution function,
	\begin{align}
	 G^<_n(\omega,t) = 2\pi i F_n(\omega,t) {\mathcal{A}}_n(\omega,t) ,\,\,\,\, \forall n \geq 1	
	\end{align}         
        \item
	Use the self-consistency Eq.~\eqref{eqn:Self_consistent_Eq_bethe} to fix the hybridization function of the effective steady state impurity model:
	\begin{align}
	\begin{cases}
	\label{Deltan1}
	\Delta_1^{R,<}(\omega,t)&=t_h^2 G^{R,<}_0(\omega,t=0)  , \,\,\,\ \text{for} \,\,\,\ n=1 \\
	\Delta_n^{R,<}(\omega,t)&=t_h^2 G^{R,<}_{n-1}(\omega,t)+\Gamma_n(\omega,t), \,\,\,\ \forall  n>1
	\end{cases}
	\end{align}
	We notice that, for $n=1$, $\Delta_1$ is proportional to $G_0$ for each time-step $t$, i.e., the impurity sees a self-consistent bath that is fixed at the initial temperature $T_i$ during the whole time propagation. Starting from $n>1$, for each timestep $t$, the impurity sees a bath which depends on the Green's function at the previous DMFT iteration $n-1$. 
    The self-energy $\Gamma_n$  that mimics photoexcitation is turned on only for global DMFT $n>0$, since at the very first iteration, $n=0$, the system is kept in the initial equilibrium state for all the times.
	\item
	Solve the impurity model. With IPT as an impurity solver, we first determine $\mathcal{G}_n(\omega)$ from $\Delta_n(\omega)$,
	\begin{align}
	 \mathcal{G}^R_n(\omega,t)&=  [\omega+i0^+ -H_{\text{loc},n}(t) - \Delta^R_n(\omega,t) ]^{-1} ,\,\,\,\, \forall n \geq 1 \nonumber\\
	\mathcal{G}^<_n(\omega,t) &= \mathcal{G}^{R}_n(\omega,t)\Delta^<_n (\omega,t)  \mathcal{G}^A_n(\omega,t) ,\,\,\,\, \forall n \geq 1
	\end{align}
     then transform to real time, evaluate Eq.~\eqref{IPT}, and transform back to frequency space to obtain $\Sigma_{U,n}^{R,<}(\omega)$.
	\item In the time-translational invariant case (NESS loop), the purely dissipative part of the self-energy, Eq.~\eqref{dephasing}, becomes
	\begin{align}
		\Sigma^{R,<}_{\gamma,\sigma,n}(\omega,t)=  \gamma_{n}(t)  \frac{1}{2 \pi} \int d \omega \ G_n^{R,<}(\omega,t)
	\end{align} 
    %\textcolor{red}{This equation should be different from the Eq.~\ref{eqn:KBE_SelfEnergy} ?}
        %\antp{With respect to Eq.~\eqref{eqn:KBE_SelfEnergy}, now $\Sigma$ and $G$ depend on the global DMFT iteration $n$, as well as $\gamma$. At the first global DMFT iteration $\gamma_{n=0}(t)=0\,\forall t$ since the system is kept in equilibrium at the initial temperature;  for $n>0$, $\gamma_n(t): 0 \to \gamma_f $ at $t=0^+$. } \textcolor{red}{In this boltzmann step by step we have the two arguments $\omega$ and $t$ for the Self-energy and green function ? }
        At the first global DMFT iteration, $n=0$, $\gamma_{n=0}(t)=0, \,\forall t$ since the system is kept in equilibrium at the initial temperature;  for $n>0$, $\gamma_n(t): 0 \to \gamma_f $ at $t=0^+$.
	\item
	Set $\Sigma_{\text{int},n}(\omega,t)=\Sigma_{U,n}(\omega,t)+\Sigma_{\gamma,\sigma,n}(\omega,t)$. 
	\item Update $F_n$ by means of QBE Eq.~\eqref{QBE}:
	\begin{align}
	\partial_t F_n ( \omega, t ) = I [F_n(\omega,t), \cdot]
	\end{align}
    \end{enumerate}
	Perform the steps 1)- 6) for the next time $t+h$ till $t_{\text{max}}$ is reached. 
	Once $t_{\text{max}}$ is reached for the iteration $n$, the new iteration $n+1$ starts.
        %from $t=0$ till $t_{max}$. 
 Convergence is reached when, for each time $t$, the functions are not updated significantly from one iteration to the other. We require
	\begin{align}
		\sum_{\omega,t} | G_n(\omega,t)-G_{n-1}(\omega,t)| <10^{-6}
	\end{align}	  
The  total energy of the system at iteration $n$ is
\begin{align}
	E_n(t)=&\frac{1}{2 \pi}\int d\omega \left \{-2i [\Delta_n(\omega,t)G_n(\omega,t)]^<\right \} \nonumber \\
    &\frac{1}{2 \pi}\int d\omega\left \{ - i [\Sigma_{U,n}(\omega,t)G_n(\omega,t)]^< \right \}
	\label{energy_QBE_n}
\end{align}

%\begin{figure*}
%    \centering
%    \includegraphics[width=1\linewidth]{figures_paper/Figure_4_t_20_beta.png}
%    \caption{Full-DMFT, quench in gamma and photoexcitation, $\beta_{eff}$.}
%    \label{fig:enter-label}
%\end{figure*}
%\begin{figure*}
%    \centering
%    \includegraphics[width=1\linewidth]{figures_paper/Figure_4_etot_t_20.png}
%    \caption{Full-DMFT, quench in gamma and photoexcitation, $E_{tot}$.}
%    \label{fig:enter-label}
%\end{figure*}

\section{Excitation protocol} 

\label{sec:ex_prot}

In order to simulate a photo-doping excitation, the system is shortly coupled with a fermionic bath with density of states
\begin{align}
    \label{Abath}
    \mathcal{A}_{{\text{bath}}}(\omega)= \mathcal{A}_b (\omega-\omega_0) + \mathcal{A}_b (\omega+\omega_0)
\end{align}
consisting of two smooth bands with bandwidth $W_{\rm bath} = 4$ centered around the energies $\omega_0 = \pm 2.5$~\cite{Picano2023_inhomogeneous}. We choose
\begin{align}
    \mathcal{A}_b (\omega) = \frac{1}{\pi} \cos^2 (\pi \omega /W_{\text{bath}})
\end{align} 
in the interval $[\omega_0-W_{\rm bath}/2,\omega_0+W_{\rm bath}/2]$.
The occupied and unoccupied density of states have spectral shapes given by $\mathcal{A}_{\rm bath}^{<} (\omega)= \mathcal{A}(\omega-\omega_0)$ and $\mathcal{A}_{\rm bath}^{>} (\omega)= \mathcal{A} (\omega+\omega_0)$, respectively. 
During the whole time-evolution of the system, the bath occupation $f_{\text{bath}}(\omega)=f_{-\beta_{\rm bath}}(\omega)$ is taken to be fixed at negative temperature Fermi-Dirac distribution (population inversion). In this way, the coupling of the system with the bath mimics photo-excitation.
In particular, $\beta_i$ corresponds to the initial (inverse) temperature of the system (in our calculations, $\beta_{\rm bath} \equiv \frac{1}{T_{\rm bath}}=20$). 

This fermionic bath adds a local contribution to the electronic self-energy $\Sigma$ in Eq.~(\ref{scatt_int}), given by
\begin{align}
    \label{Gamma}
    \Gamma^{<(>)} (\omega,t) &=(-)2\pi i V^2(t) \mathcal{A}_{\text{bath}}^{<(>)} (\omega) \\
    \Gamma^R(\omega,t) &= - i \pi V^2(t) \mathcal{A}_{\text{bath}}(\omega)
    \label{Gamma_ret}
\end{align}
with time-dependent profile 
\begin{align}
    \label{V}
    V(t) =  \Gamma \sin^2 (\pi t/T_0) \theta (t) \theta (T_0-t),
\end{align}
where $T_0=1$ (the pulse duration) in units of hopping times and $\Gamma$ is the amplitude of the coupling between the system and the bath.
%We choose $A/t_h=0.1$, $A/t_h=0.5$, and $A/t_h=1$ (see for example Fig.~\ref{fig:QuenchPhotoexcitation} in the main text). 
%\textcolor{red}{Should we explain a bit the physical interpretation of all this parameters ? This type photoexcitation is not commun in the open quantum system field no ? in the main text or in the appendix } \antp{yes, I will explain it in blue and give some context about the photoexcitation protocol.}
In terms of the two times $(t,t')$ on the Keldysh contour $\mathcal{C}$, the bath reads
\begin{align}
	\label{sphonto}
	\Gamma(t,t')=V(t)G_{\text{bath}}(t,t')V(t')^*,
\end{align}
where the bath Green's function $G_{\text{bath}}(t,t')$ is defined as
\begin{align}
	G_{\text{bath}}^R(t,t')
	&=
	-i
	\theta(t-t')
	\int d\omega \,e^{-i\omega(t-t')}\mathcal{A}_{\text{bath}}(\omega),
	\\
	G_{\text{bath}}^<(t,t')
	&=
	i
	\int d\omega\,
	e^{-i\omega(t-t')} f_{\text{bath}}(\omega)\mathcal{A}_{\text{bath}}(\omega)
\end{align}

\section{Emergence of Infinite Temperature Thermalization: steady-state properties. }
\label{sec:beta_n}

In this section, we present additional results illustrating the emergence of the infinite-temperature state. As discussed in Sec.~\ref{sec:front}, the dynamics observed within the DMFT self-consistency loop deviate from those of purely unitary evolution: the effective temperature does not return to its initial value. Instead, the system evolves toward a prethermal steady state at long times, signaling the onset of thermalization driven by dissipation.

In Fig.~\ref{fig:prethermal_steadystate_beta}, we show the inverse temperature $\beta_{\rm eff,n}$ as a function of the step-by-step DMFT iteration $n$, for different values of the dissipation strength $\gamma$.
%\textcolor{red}{Add some text about the figure}
For large $n$, the effective temperature becomes constant as a function of $n$.
This means that the step-by-step solution has converged to the full-DMFT one.
As the dissipation strength increases,
the final constant value of the effective temperature increases, since the system gets closer to the infinite temperature limit.  (For $\gamma=0.6$ we don't see the plateau only because it happens for $\beta \approx 10^{-4}$ for $U=0$, and $\beta \approx 10^{-5}$ for $U=2$ ). The way the effective temperature approaches this plateau depends on the value of the Hubbard interaction $U$ (and $\gamma$). For $U=0$ (Fig~\ref{fig:prethermal_steadystate_beta} (a)),  the 
decay is exponential with $n$ for all $\gamma$. 
For $U=2$ (Fig~\ref{fig:prethermal_steadystate_beta} (b)),
it shows a double-exponential decay for $\gamma=0.6$. for lower values of $\gamma$, the system spends more and more time at equilibrium at the initial temperature $\beta_i=20$, before relaxing to the final state with a non-trivial dynamics.

%\section{Emergence of Infinite Temperature Thermalization in the Total Energy}
%In this section, we present additional results on the emergence of the infinite-temperature state, focusing on the behavior of the total energy $E_{\rm tot}$. In Fig.~\ref{fig:step-by-step-quench-gamma-photoexc}, we plot the dynamics of the total energy at different DMFT iteration steps and for increasing values of dephasing rate $\gamma$ at fixed photo-excitation amplitude $A/t_h = 1$.

%Interestingly, the dynamics of the total energy reveal that when $\gamma \ll A$, the thermalization front is nearly linear in time. At intermediate values of $\gamma$, the front begins to exhibit a parabolic shape, and when the dephasing rate becomes comparable to the photoexcitation amplitude $\gamma \sim A$, the thermalization front disappears entirely, meaning that the DMFT loop converge in one iteration. 

%\textcolor{red}{A finir}

%\bibliography{heating_ref} 

%

\end{document}